\documentclass[journal=langd5,manuscript=article]{achemso}

\usepackage[version=3]{mhchem} 

\usepackage{graphicx}

\usepackage{amsmath}
\usepackage{amssymb}

\usepackage{bm}

\usepackage{ifpdf}

\usepackage{caption}
\usepackage{subcaption}

\usepackage{color}
\usepackage[usenames,dvipsnames]{xcolor}

\newcommand*{\citen}[1]{
  \begingroup
    \romannumeral-`\x 
    \setcitestyle{numbers}%
    \cite{#1}%
  \endgroup   
}

\DeclareMathOperator*{\define}{\equiv}

\newcommand{\eq}[1]{ Eq.\ (\ref{#1})}

\newcommand{\figwidth}{0.7}

\newcommand{\eg}{\textit{e.g., }}
\newcommand{\ie}{\textit{i.e., }}

\newcommand*{\xlineshort}[1][1.0em]{\rule[0.4ex]{3.5pt}{0.5pt}}

\newcommand*{\xdash}[1][1.0em]{\rule[0.01ex]{2.5pt}{0.5pt} \rule[0.01ex]{0.5pt}{0.5pt}}
\newcommand*{\xdashthick}[1][1.0em]{\rule[0.5ex]{2.5pt}{1.5pt} \; \rule[0.5ex]{2.5pt}{1.5pt}}
\newcommand*{\xdashvthick}[1][1.0em]{\rule[0.4ex]{4.5pt}{2.5pt} \rule[0.4ex]{4.5pt}{2.5pt}}

\definecolor{lllgrey}{rgb}{0.9,0.9,0.9}
\definecolor{llgrey}{rgb}{0.8,0.8,0.8}
\definecolor{lgrey}{rgb}{0.6,0.6,0.6}
\definecolor{dlgrey}{rgb}{0.4,0.4,0.4}
\definecolor{ddlgrey}{rgb}{0.2,0.2,0.2}
\definecolor{cm_orange}{rgb}{0.9950,0.8090,0.5000}
\definecolor{cm_red}{rgb}{0.9520,0.4180,0.2580}
\definecolor{cm_blue}{rgb}{0.3680,0.5750,0.7660}

\definecolor{fft_darkblue}{rgb}{0.19215686619281769, 0.21176470816135406, 0.58431375026702881}
\definecolor{fft_lightblue}{rgb}{0.56485968018118948, 0.7663975529283894, 0.86758939636894861}
\definecolor{fft_yellow}{rgb}{0.99992310649750094, 0.99761630142252977, 0.74540562818527956}
\definecolor{fft_lightred}{rgb}{0.97347174068223274, 0.54740483716834987, 0.31810842222408564}
\definecolor{fft_darkred}{rgb}{0.64705884456634521, 0.0, 0.14901961386203766}

\newcommand*\chem[1]{\ensuremath{\mathrm{#1}}}

\author{Edward R. Smith}
\affiliation{Department of Civil Engineering, Imperial College London, South Kensington Campus, London SW7 2AZ, United Kingdom}
\email{edward.smith05@imperial.ac.uk}
\author{Panagiotis E. Theodorakis}
\affiliation{Institute of Physics, Polish Academy of Sciences, Al. Lotnik\'ow 32/46, 02-668 Warsaw, Poland}
\email{panos@ifpan.edu.pl}
\author{Richard V. Craster}
\affiliation{Department of Mathematics, Imperial College London, South Kensington Campus, London SW7 2AZ, United Kingdom}
\email{r.craster@imperial.ac.uk}
\author{Omar K. Matar}
\affiliation{Department of Chemical Engineering, Imperial College London, South Kensington Campus, London SW7 2AZ, United Kingdom}
\email{o.matar@imperial.ac.uk}

\title{Moving contact lines: linking molecular dynamics and continuum-scale modelling}

\abbreviations{}

\begin{document}

\begin{abstract}
Despite decades of research, the modelling of moving contact lines has remained a formidable challenge in fluid dynamics whose resolution will impact numerous industrial, biological, and daily-life applications. 
On the one hand, molecular dynamics (MD) simulation has the ability to provide unique insight into the microscopic details that determine the dynamic behaviour of the contact line, which is not possible with either continuum-scale simulations or experiments. On the other hand, continuum-based models provide the link to the macroscopic description of the system. In this Feature Article, we explore the complex range of physical factors, including the presence of surfactants, which govern the contact line motion through MD simulations. We also discuss links between continuum- and molecular-scale modelling, and highlight the opportunities for future developments in this area. 







%
%
%

\end{abstract}

\maketitle

\section{Introduction}
\label{Sec:Introduction}

The nature of matter has long been a source of philosophical debate:  
Around 400BC, Democritus and Leucippus postulated an indivisible unit which they called the atom \citep{Coveney_highfield}.
With no evidence for the existence of atoms and the apparent continuous nature of matter, 
the discrete paradigm largely fell from favor over the ensuing millennia. 
This view was so ingrained by the late 19$^{\rm th}$ century that, despite being central to Ludwig Boltzmann's development
of statistical mechanics, discrete atoms were regarded as little more than a tool. 
As a result of this dichotomy, the fields of continuum mechanics, and discrete particle dynamics evolved separately, 
well into the age of computers. Even today, the two descriptions are often studied independently 
by separate research communities. 

\begin{figure}
\includegraphics[width=\figwidth\textwidth]{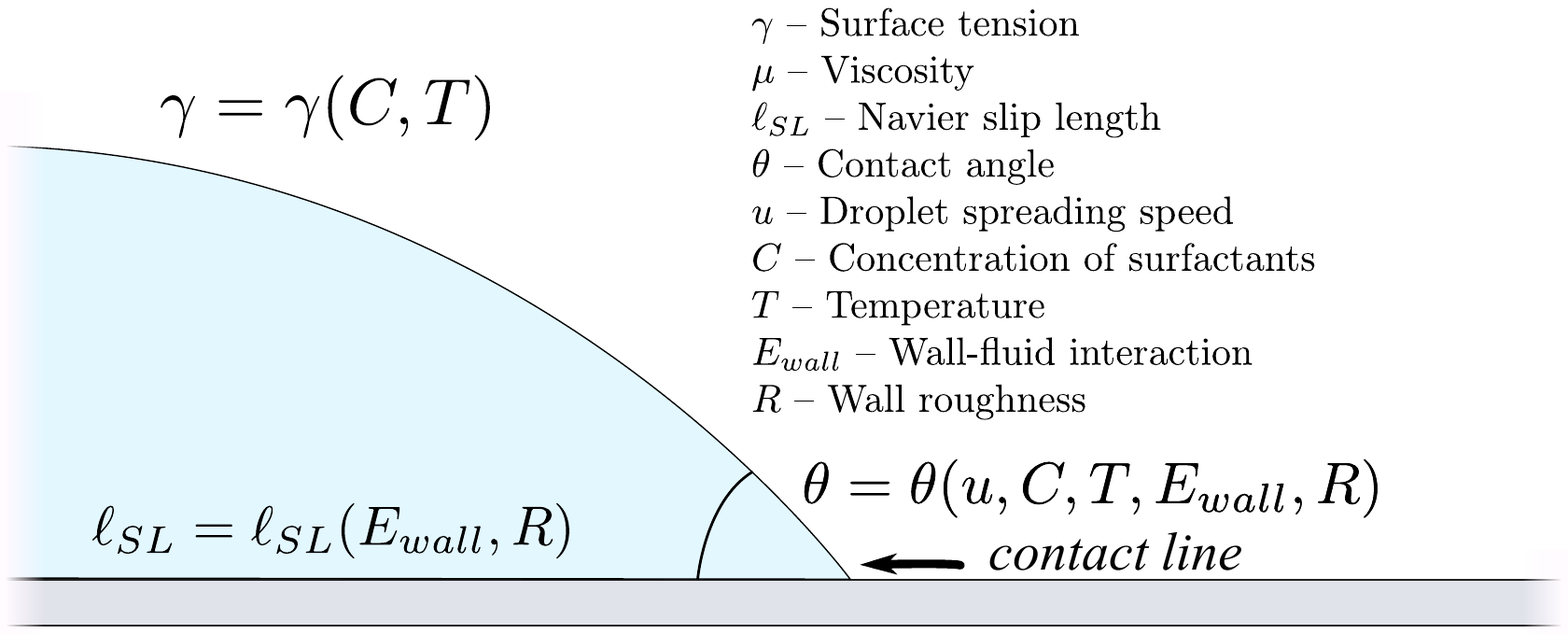}
\caption{Schematic representation of a contact line for a sessile drop on a solid substrate. 
} 
\label{contact_angle_schematic}
\end{figure}

The {\it continuum hypothesis} is one of the primary cornerstones of predictive models for many diverse engineering applications.  There are instances, however, when it becomes necessary to explore motion at the meso- and molecular-scales for which this hypothesis requires refinement. 
A prime example of this is the modelling of moving contact lines \citep{Snoeijer_Andreotti}, which is a feature of many flows where an interface separating two phases intersects a solid boundary, and moves despite the classical requirement for imposing the no-slip condition.\cite{deGennes1985} 

Understanding the behaviour of the contact line is very important for many applications including a recent exciting area, that is spontaneous dewetting of ultra thin films.\cite{Mukherjee2015}
To model such flows, one actually needs to combine three separate problems: $1)$ capture the behaviour of the system at the interfaces between the solid substrate and the liquid, $2)$ model the interaction of the solid substrate and the air, as well as $3)$ reproduce the dynamic interface between the liquid surface and the air (see Fig. \ref{contact_angle_schematic}).
The meeting point of all three problems is at the so called {\it contact line}. 
Continuum approaches at the contact line are unable to elucidate the microscopic underlying mechanisms, which involves understanding the diffuse and complex nature of the contact line.
In contrast to molecular-level models \citep{Razavi2014}, the behaviour of the contact line is an assumption of a continuum-based model and not a result of the numerical simulation.\cite{Karapetas_et_al,Beacham2009,Karapetsas2011b} 
As a result, the dynamics of the contact line should be addressed at the molecular level.

Molecular dynamics (MD) has demonstrated
that although a microscopic resolution at the contact line is indispensable for understanding wetting-type phenomena, it is often also important to take into account areas of the liquid far from the interfaces or the contact line in order to provide a complete description of the system in tandem with the contact line. 
For example, in the presence of additional complexities associated, for instance, with surfactant-laden systems  \cite{Theodorakis2014,Venzmer2011,Nikolov2011} 
the bulk of the droplet acts as a source of surfactants for the interfaces, which subsequently supplies the contact line with surfactants.\cite{Theodorakis2015,Theodorakis2015b} 
Although a molecular-level description of the surfactants in the bulk of the droplet may be unnecessary, the dynamics of the system is determined by the interplay between the surfactant in the bulk and the interfaces/contact line.\cite{Theodorakis2015,Theodorakis2015b} 
In this case, the dynamic behaviour of the contact line often requires a microscopic resolution, whereas the bulk can be described by continuum-scale models.
Using simple molecular models, which can predict the contact angle, is also possible. For example, in the case of a liquid bridge sandwiched between parallel plates, which will be discussed later, one can demonstrate good agreement in terms of the meniscus shape and the static angles, \cite{Thompson_et_al89,Thompson_et_al93} and contact line dynamics can be explored \citep{Smith2016}.

In this Feature Article, we give our perspective on the contact line problem, in which we classify and consider the various approaches for linking molecular- and continuum-level models. We start with direct coupling of an MD and a Computational Fluid Dynamics (CFD) solver to model near-wall dynamics, before moving on to parameterization of quantities such as pressure, surface tension, static and dynamic contact angles.
Direct coupling allows the molecular and continuum regions to evolve together as parts of one simulation, while parameterizations allow molecular details to be abstracted and included in a CFD solver which will model the whole space.
We present the role of near-wall interactions, and discuss the dynamic and static angle behaviour in the context of two flow configurations: $1)$ a liquid bridge as mentioned above, undergoing shear flow; and $2)$ a droplet spreading in the presence and absence of surfactants. 
The results from a sheared liquid bridge, a commonly used approach to study the contact line in the MD literature, are discussed critically motivating the final simulation of a full MD droplet. 
The work is organized as follows: 
In the Methodology  Section, \ref{Sec:Methodology} we give the outline of our continuum and coarse-grained (CG) MD models. In the Results and Discussion Section, \ref{sec:results} we discuss  results for the contact line dynamics; in particular, the near-wall interaction, surface tension,  static and dynamic contact angles,  electrowetting, droplet modelling, and the link between molecular models and continuum-based approaches.  
We conclude our article by highlighting limitations of our methods, the importance of accessing larger systems with microscopic resolution, and future directions.

\section{Methodology}
\label{Sec:Methodology}

In this section, we outline the continuum and molecular modelling methodologies used in this work, and highlight the links between them.

\subsection{Continuum dynamics}

Continuum-scale theories rely on conservation principles to provide tools for predicting the spatio-temporal evolution of `fields', such as the fluid velocity, pressure, and, for non-isothermal flows, temperature. An equation for the velocity ${\mathbf{u}}$ is provided by the equation of force-momentum balance as described in, say, Batchelor\cite{batchelor67a}. Here, the rate of change of momentum is determined by momentum advection, and the balance of forces over an arbitrary volume, $V$, enclosed by a surface with area $S$:
\begin{align}
	\!\! \frac{d }{d t}  \int_V  \rho \mathbf{u} dV \! = -
\oint_S \left[ \rho \mathbf{u}:\mathbf{u} + \boldsymbol{\Pi} \right]
\cdot d\textbf{S}  + \textbf{F}_{\textnormal{body}},
\label{BofMEqn2}
\end{align}where $\rho$ is the density of the fluid, $\mathbf{u}$ is the fluid velocity, $\boldsymbol{\Pi}$ 
the stress 
tensor, $\textbf{F}_{\textnormal{body}}$ denotes external forces (per unit volume) acting on the fluid, and $t$ represents time. 
By assuming that the continuum hypothesis is valid, we take the zero volume limit in the momentum balance, \eq{BofMEqn2}, to arrive at the celebrated Navier--Stokes equations:  

\begin{align}
	\rho\left(\frac{\partial \mathbf{u}}{\partial t} + \left( \mathbf{u} \cdot \mathbf{\nabla} \right) {\mathbf u}\right)
					= -  \mathbf{\nabla} P -\rho {\mathbf{g}}+ \mu\nabla^2 \mathbf{u}. 
\label{ANDNSEqn}
\end{align}Here, we have assumed that $\mathbf{\Pi}=-P{\mathbf{I}} + \mu\left(\nabla {\mathbf{u}}+\nabla{\mathbf{u}}^{T}\right)$ in which $\mu$ is the fluid (constant) viscosity, $P$ is the fluid pressure  
and $\mathbf{I}$ is the identity tensor. 
\citep{Gad-El-Hak_06} 
We have also set ${\mathbf{F}}_{\textnormal{body}}=-\rho {\mathbf{g}}$ reflecting the presence of gravitational forces in many applications, though these will be neglected in the systems which are considered in the present work.  
In addition to the momentum balance, an equation of mass conservation must also be solved, given by 
\begin{align}
\boldsymbol{\nabla}\cdot \mathbf{u} = 0;
\label{incompressible_cont}
\end{align}
which is consistent with the assumption of an incompressible fluid. 

Computational Fluid Dynamics aims to solve \eq{ANDNSEqn} and \eq{incompressible_cont} using numerical approximations for the derivatives in $\mathbf{u}$ and $P$. 
Numerical solutions of these equations are obtained starting from appropriate initial conditions, which reflect the particular physical situation under consideration. These solutions are subject to boundary conditions on $\mathbf{u}$, which correspond to the so-called `no-slip', and `no-penetration' conditions on the tangential and wall-normal components of the velocity, respectively. The  no-penetration condition is appropriate for situations in which the solid substrate underlying the fluid is impermeable. The no-slip condition, in turn, is commonly deployed in, for instance, single-phase, Newtonian flows in pipes and channels. This condition, however, must be modified when modelling systems involving, for instance, the spreading of a droplet on a solid substrate; otherwise, the no-slip condition leads naturally to a stress singularity at the moving contact line \citep{Bonn2009}. 
%

Instead, a 
Navier-slip model can be used \citep{Navier,Bonn2009}, given by the following expression
\begin{align}
 u = \ell_{sl} \frac{\partial u}{\partial z}, 
 \label{Navier_slip}
\end{align}
where $\ell_{sl}$ is the slip length, $u$ denotes the streamwise component of $\mathbf{u}$, and $z$ the wall-normal coordinate; setting $\ell_{sl}=0$ recovers the no-slip condition.  

The lubrication approximation is often used to model the spreading of slender droplets via 
solution of
a reduced form of Eqs. (\ref{ANDNSEqn}) and (\ref{incompressible_cont})  \citep{Karapetas_et_al}: 
\begin{align}
 \frac{\partial u}{\partial x}  + \frac{\partial w}{\partial z} = 0  ; \;\;\;  \frac{\partial P}{\partial x} = \frac{\partial^2 u}{\partial z^2} ; \;\;\; \frac{\partial P}{\partial z} = 0,
\label{thin_film}
\end{align}
where $x$ represents the streamwise direction, and $w$ denotes the velocity component in the wall-normal direction; here, gravitational forces have been neglected.
Solutions of these reduced equations are subject to the kinematic, normal, and tangential stress conditions at the interface, $z=h$, respectively given by
\begin{align}
 \frac{\partial h}{\partial t} + u \frac{\partial h}{\partial x} = w ; \;\;\;  P = P_0-  \gamma \frac{\partial^2 h}{\partial x^2} ; \;\;\; \mu\frac{\partial u}{\partial z}=\frac{\partial \gamma}{\partial x}, &  &\text{ at }  z = h,
\end{align}
and the Navier-slip condition, \eq{Navier_slip}, for the wall boundary at $z=0$. In the normal-stress condition, $P_0$ denotes the ambient pressure (which can be set to zero without loss of generality), and the second term on the right-hand-side represents capillary effects: surface tension, $\gamma$, multiplying the interfacial curvature (in its simplified form in the lubrication approximation). In the tangential-stress condition, we have included on the right-hand-side the possibility of surface tension gradients, which arise, for instance, in the presence of surfactant where $\gamma$ varies with surfactant concentration. 
The thin film equation is introduced here as a simple model for spreading droplets, however in practice a more comprehensive fluid solver would be required for the general coupling presented in this work.

A closure is required for the contact line speed $u_{cl}$, the simplest and most common form of which is represented by Tanner's law \citep{Tanners}:
\begin{align}
u_{cl} = A_{cl} \left( \langle \theta \rangle - \theta_e \right)^n,
\label{Tanners}
\end{align}
where $\langle \theta \rangle$ is the mean local angle,  $\theta_e$ is the equilibrium contact angle, and $A_{cl}$ and $n$ are constants. 
It is worthwhile noting that the continuum-scale modelling is well established with review articles  \cite{Blake_2006,Bonn2009,Craster2009,Sui_2014} covering many of the aspects briefly discussed above.
We note here that Tanner's law involves the apparent contact angle, which may be different than the microscopic contact
angle, defined at the solid surface.\cite{Bonn2009} The latter requires averaging over 
microscopic data\cite{Thompson_Robbins} or the application of well-defined theoretical models that take into account
the diffuse nature of the interface.\cite{Qian2003} 

In what follows, we describe methodologies used to provide estimates for $\langle\theta\rangle$ from averaged values of the contact angle obtained via molecular dynamics simulations.


\subsection{Molecular Dynamics}

Molecular modelling is a direct solution of Newton's Laws for individual molecules and shows excellent agreement to experiments, reproducing the underlying liquid structure (matched to x ray scattering \citep{Rapaport}); equilibrium thermodynamic properties (phase diagrams, triple point \citep{sadus_13, sadus_16}); flow dynamics and diffusion \citep{Alder_Wainwright70}; \textit{a priori} prediction of dynamic coefficients such as viscosity, surface tension \citep{Smith_et_al2016}, heat flux \citep{Todd_Evans_95} and slip-lengths \citep{Thompson1997, Hansen_et_al2011}; as well as fluid dynamics for canonical flows like Couette \citep{Smith_et_al}, Poiseuille \citep{Travis_Gubbins_00}, non-linear flow case like Rayleigh-B\'{e}rnard instability \citep{Rapaport_88}, complex chemical superspreading \citep{Theodorakis2015b}, shock waves dynamics \citep{Hoover_1979, Root_et_al} and even turbulence \citep{Smith_2015}. 
Naturally, high resolution all-atom simulations is therefore an ideal tool to study the dynamics of the contact line. Yet, phenomena in this region of a fluid can involve physics at larger length- and time-scales than is practicable to model atomistically. \cite{Halverson2009} 
In addition, all-atom force-fields require a parametrization, which, in turn, involve targeting the reproducibility of certain  properties of a system (e.g. thermophysical ones)\cite{Vega2007} or, in the case of complex molecules (e.g. proteins), 
their native structures\cite{Poma2017}. 
To overcome these constraints, coarse-grained (CG) force-fields, that provide adequate resolution for the phenomena involved at the contact line, but at the same time allow access to large time and length scales\cite{Marrink2007,Sergi2012} have been developed. 

A popular class of models, Statistical Associating Fluid Theory  (SAFT) initiated in the late 1980s \cite{Chapman1989}, with their progress at the turn of the century reviewed in Ref.~\citen{Muller2001}, have been refined to a level where they provide a realistic methodology sophisticated enough to capture the essential thermophysical behaviours observed. The SAFT-$\gamma$ we use here is a molecular-based equation of state 
\cite{Lafitte2013,Avendano2011,Avendano2013,Muller2014,Papaioannou2014} 
which offers an accurate fit for the force-field parameters that can be optimized to reproduce the macroscopically-observed thermophysical properties.\cite{Herdes2015,Lobanova_PhD,Lobanova2015,Muller2014,Avendano2011,Avendano2013}
Hence, SAFT-$\gamma$ ensures that the potential parameters are consistent with experimental results on thermodynamic data 
(first and second derivatives of the free energy) while retaining much of the simplicity of Lennard--Jones (LJ) systems,
as interactions between groups of atoms are modelled with the Mie potential,  a generalized LJ potential that offers 
larger flexibility in the fitting of the equation of state. As a result, problem complexity does not obscure the relevant physical mechanisms, which are a central motivation for molecular simulation.

SAFT-$\gamma$ has provided the potential parameters for different components of the CG system including those for
water.\cite{Lobanova2015,Theodorakis2015,Theodorakis2015b} In the case of the SAFT-$\gamma$ water model we use here, 
one spherical bead in the simulations represents two water molecules allowing for the simulation of even larger systems.  
The model has been tested to reproduce thermodynamic data and other properties, such as the surface tension of water.\cite{Lobanova_PhD,Lobanova2015}
Based on this CG model, the range of problems that can be addressed also include systems involving water molecules and
surfactants.\cite{Theodorakis2015,Theodorakis2015b,Lobanova_PhD,Lobanova2015} 

In the molecular model, the forces exerted on groups of atoms represented by effective beads  
are $\boldsymbol{f}_{\rm ij} = -\boldsymbol{\nabla} \phi_{\rm ij}$, where $\phi_{\rm ij}$ is the Mie potential
between beads of type $\rm i$ and $\rm j$ being a 
distance $r_{\rm ij}$ apart:
\begin{align}
  \phi_{\rm ij} = C \epsilon_{\rm ij} \left[ \left( \frac{\sigma_{\rm ij}}{r_{\rm ij}} \right)^{\lambda_{\rm ij}^r} - \left( \frac{\sigma_{\rm ij}}{r_{\rm ij}} \right)^{\lambda_{\rm ij}^a} \right],
  \label{Mie_potential}
\end{align}
and $C$ is given by,
 \begin{align}
C = \left( \frac{\lambda_{\rm ij}^r}{\lambda_{\rm ij}^r-\lambda_{\rm ij}^a}\right) \left( \frac{\lambda_{\rm ij}^r}{\lambda_{\rm ij}^a}\right)^{\left( \frac{\lambda_{\rm ij}^a}{\lambda_{\rm ij}^r-\lambda_{\rm ij}^a}\right)}. 
\end{align}
The parameters $\lambda_{\rm ij}^a$ and $\lambda_{\rm ij}^r$ are defined for different pairs of interacting beads $\rm i$ 
and $\rm j$. To this end, the exponent $\lambda_{\rm ij}^a$ has only a physical meaning
expressing the dispersion interaction between different beads, whereas $\lambda_{\rm ij}^r$ acts as a fitting parameter and affects the core interactions between beads (size of beads).
In the CG model, the unit of length is $\sigma$ and that of energy $\epsilon$, 
which correspond to $\sigma = 0.436$ nm and $\epsilon/k_B = 492$ K, respectively, and $k_B$ is the Boltzmann constant. 
The cross interaction parameters are predicted by the following combination rules \cite{Lafitte_et_al}
\begin{align}
\sigma_{\rm ij} = 0.5\left[\sigma_{\rm ii} + \sigma_{\rm jj}\right], \;\;\;\;\;\;
\epsilon_{\rm ij} = \frac{\sqrt{\sigma_{\rm ii}^3 + \sigma_{\rm jj}^3}}{\sigma_{\rm ij}^3} \sqrt{\epsilon_{\rm ii}\epsilon_{\rm jj}}
\end{align}
with
\begin{align}
\lambda_{\rm ij}^k - 3 = \sqrt{(\lambda_{\rm ii}^k - 3)(\lambda_{\rm jj}^k - 3)}, \;\;\; k = a,r.
\end{align}
Unless stated otherwise, the model assumes a universal cutoff, which is $r_c=4.5834\sigma$.
The values of the potential parameters for different groups of atoms represented by the effective beads Ar, W, M, D, EO, and CM are given in Table \ref{SAFT_table}; $\lambda_{\rm ij}^a = 6$ for all effective beads. Here, one Ar atom is represented by a single bead. The following atoms are also represented by a single bead as follows: 
W for two water molecules [$\chem{H_2O}$],  M and D for $\left[\chem{(CH_3)_3-Si-O_{\frac{1}{2}}}\right]$ and 
$\left[\chem{O_{\frac{1}{2}}-(CH_3)_2-Si-O_{\frac{1}{2}}}\right]$, respectively,  EO for  $\left[\chem{-CH_2-O-CH_2-}\right]$ 
as well as CM for $\left[\chem{-CH_2-CH_2-CH_2-}\right]$. 

\begin{center}
    \captionof{table}{Potential parameter values for different effective beads representing different 
    groups of atoms as discussed in the text. In parenthesis, we
    provide the units for each parameter.}
	\begin{tabular}{|c|c|c|c|c|}\hline
		Mol & Mass (m) & $\sigma_{\rm ii} (\sigma)$ & $\epsilon_{\rm ii} (\epsilon)$ & $\lambda_{\rm ii}^r$\\\hline
		Ar & 0.90645 & 0.782 & 0.2429 & 12.0 \\\hline
		W & 0.8179 & 0.8584 & 0.8129 & 8.0 \\\hline
		M & 1.8588 & 1.2398 & 0.8998 & 26.0 \\\hline
		D & 1.6833 & 1.6833 & 0.5081 & 13.9 \\\hline
		EO & 1.0000 & 0.9307 & 0.8067 & 19.0 \\\hline
		CM & 0.9552 & 1.0000 & 0.7000 & 15.0 \\\hline
	\end{tabular}
	\label{SAFT_table}
\end{center} 

If one is interested in accounting for the presence of surfactant molecules, then with the EO, D, and CM one can build CG models for both so-called ``superspreading" and 
non-superspreading surfactants.\cite{Theodorakis2015}
In the case of surfactants, chains are built by binding effective beads with a harmonic potential \cite{Lobanova2015,Lobanova_PhD}: 
\begin{equation}
\label{harmonicbond}
\phi_B(r_{\rm ij}) = 0.5 k (r_{\rm ij}-\sigma_{\rm ij})^2,
\end{equation}
where 
the values of $\sigma_{\rm ij}$ are given in Table \ref{SAFT_table},   
and $k=295.33 \epsilon/\sigma^2$. Additionally, any three consecutive beads in a surfactant molecule
of type EO interact via a harmonic angle potential
\begin{equation}
\label{harmonicangle}
\phi_{\theta}(\theta_{\rm ijk}) = 0.5 k_{\theta} (\theta_{\rm ijk}-\theta_0)^2,
\end{equation}
where $\theta_{\rm ijk}$ is the angle defined by three consecutive beads along 
the surfactant chain, $k_{\theta}=4.32 \varepsilon/{\rm rad}^2$ is a constant, and $\theta_0=2.75$ rad is the equilibrium angle of the harmonic potential.

In an MD simulation, one typically integrates Newton's equations of motion with periodic boundaries. 
To model walls, the molecules are tethered to their equilibrium lattice sites by harmonic interactions governed by relations such as that expressed by Eq. (\ref{harmonicbond}). 
In this case, a  Nos\'{e}--Hoover thermostat is applied to the tethered molecules only, and the equations of motion for the wall atoms are given by,
\begin{subequations}
\begin{eqnarray}
\boldsymbol{v}_{i}  &=& \frac{{\boldsymbol{p}}_{i}}{m_i} + U_\text{w} \textbf{n}_x, \\
\dot{{\boldsymbol{p}}}_{\rm i}  &=& \boldsymbol{F}_{\rm i} + \boldsymbol{F}_{\rm i_{\textnormal{teth}}} - \xi {\boldsymbol{p}}_{\rm i}, \\
\dot{\xi} &=& \frac{1}{Q_{\rm \xi}} \left[  \displaystyle\sum_{n=1}^{N} \frac{{\boldsymbol{p}}_{\rm n} \cdot {\boldsymbol{p}}_{\rm n}}{m_{\rm n}} -3T_{\rm 0} \right], \\
\boldsymbol{F}_{\rm i_{\textnormal{teth}}} &=& \boldsymbol{r}_{\rm i_0}  \left( 4 k_{4} r_{\rm i_0}^{\rm 2}+6 k_{6} r_{\rm i_0}^{4} \right).
 \label{NH_verify} 
\end{eqnarray}
\end{subequations}
Use of a wall-only thermostat allows a temperature profile to develop in the domain, prevents the thermostat from impacting the dynamics of the fluid \citep{Smith2016}, and represents more closely experimental setups. 
Here $\boldsymbol{p}_{\rm i} $ $\define \boldsymbol{v}_{\rm i} - U_{\rm w} \textbf{n}_x$ is the peculiar momentum, defined as the particle velocity $\boldsymbol{v}_{\rm i}$ (times particle mass $m_{\rm i}$) minus average streaming velocity, which is the wall velocity $U_{\rm w}$ in the $x$ direction denoted by vector $\textbf{n}_x$.
The molecules are tethered to equilibrium location $\boldsymbol{r}_0$ and experience a force $\boldsymbol{F}_{\rm i_{\textnormal{teth}}}$ proportional to displacement from this site, $\boldsymbol{r}_{\rm i0} \define \boldsymbol{r}_{\rm i} - \boldsymbol{r}_{\rm 0}$, with spring coefficients $k_4=5\times10^3$ and $k_6=5\times10^6$ from \citet{Petravic_Harrowell}.
Both the molecule and its tethering site slides with speed $U_w$.
The arbitrary wall thermostatting coefficient $Q_{\rm \xi}$ is chosen to be equal to $0.1 N_{_{\rm thermo}} \Delta t$ so the heat bath is proportional to system size and timestep in guiding the system to thermostat setpoint $T_{\rm 0}$.
Our simulations have been performed using an in-house MD code, called {\it flowmol} \citep{Smith_Thesis}. 

A more simplified way of considering solid substrates in MD simulations is an 
unstructured smooth wall,\cite{Theodorakis2015,Theodorakis2015b} which is realized by using an interaction 
potential that depends on the distance between the effective beads and the wall.\cite{Israelachvili,Forte2014} In this case,
the fluid--substrate interactions can be modelled by an unbiased integration (i.e. density inhomogeneities and structural characteristics 
of the substrate at the microscopic level are neglected)
of the solid potential considering a wall composed of spherical Mie beads, where the width of the substrate exceeds 
the cut-off of the potential.\cite{Forte2014} The form of the potential reads
\begin{equation}
\label{eq:sub}
\phi_{\rm sub}(D) = 2 \pi\rho_{\rm MD} C \epsilon_{\rm ij}\sigma_{\rm ij}^3
\left[ A  \left( \frac{\sigma_{\rm ij}}{D} \right)^{\lambda_{\rm ij}^r-3} 
- B \left( \frac{\sigma_{\rm ij}}{D} \right)^{\lambda_{\rm ij}^a-3} \right], 
\end{equation}
where $A=1/(\lambda_{\rm ij}^r-2)(\lambda_{\rm ij}^r-3)$ and $B=1/(\lambda_{\rm ij}^a-2)(\lambda_{\rm ij}^a-3)$.
$C$, $\sigma_{\rm ij}$,$\epsilon_{\rm ij}$, $\lambda_{\rm ij}^r$, and $\lambda_{\rm ij}^a$ have been defined in \eq{Mie_potential}, 
$\rho_{\rm MD}$ is the number density, which typically for a paraffinic substrate is $\rho_{\rm MD}\approx 1\sigma^{-3}$. $D$ is the
vertical distance between beads and the substrate (wall). The cut-off of the fluid--substrate interaction is the same as
the cut-off used for the fluid--fluid interactions.

In this model, the substrate--fluid interaction is tuned against the contact angle of water.\cite{Theodorakis2015b} For example, a contact angle of approximately $60^\circ$ is obtained by 
setting the value of $\epsilon_{ \rm SW}=1.4 \epsilon$, where $\epsilon_{ \rm SW}$ is the strength of interaction between the substrate and water effective beads. Then, by knowing the interaction parameter between water molecules 
(fluid--fluid interactions from Table \ref{SAFT_table}) we can obtain an estimate of an effective interaction parameter $\epsilon_{\rm SS}$ for the substrate beads by using the
above combination rules. All other fluid--solid interactions arise from the use of these combination rules. Patterned substrates of different geometry can
also be simulated using MD methods, for example in the context of
realizing equal-sized mesoscale polymer droplets of two constituent polymers by
sequential spin dewetting.\cite{Bhandaru2014} For this type of systems, MD is particularly
suitable, because one can accurately design any substrate geometry. Hence, the substrate
pattern can be constructed according to the application under consideration.\cite{Bhandaru2014}
We turn our attention now to the strategies employed for coupling molecular and continuum-scale models.

\subsection{Coupling of molecular and continuum models}
\label{sec:coupling}

There are, broadly speaking, three techniques, which we term `Types 1, 2 and 3', for linking the continuum and molecular models described in the previous sections:
\begin{enumerate}
 \item Type 1: run MD simulation to get tables of data \citep{sadus_13, sadus_16}, or a reduced model to include in the continuum simulation, such as surface tension in surfactant-laden flows or the contact line dynamics. This type of simulation includes the wider class of parameterizing continuum constants using molecular dynamics \citep{Evans_Morris}, defining non-local viscosity kernels \citep{Hansen_et_al}, defining slip boundary conditions \citep{Hansen_et_al2011, Qian2003,Qian2004,Qian2006} or any parameterization which allows the continuum model to be run for the whole space with no further MD simulation.
This type of coupling is the focus in this work, to get surface tension values, define contact angles and create a model for the moving contact line which includes fluctuations. \citep{Smith_et_al2016}
  \item Type 2: call dynamically or spawn new representative MD simulations during a continuum run to obtain (or check) parameters which are transferred to the continuum run, including effect of complex molecules on viscosity \citep{Yamamoto, E_et_al}, or slip-length. \citep{Asproulis2013}
 \item Type 3: directly link molecular and continuum solvers with each solving a portion of the same domain with mass, momentum, and energy exchange at the interface and both models evolving together \citep{ OConnell_Thompson, Li_et_al, Hadjiconstantinou_thesis, Flekkoy_et_al, Wagner_et_al, Delgado-Buscalioni_Coveney_03, Nie_et_al, Werder_et_al, Borg_et_al, Smith_Thesis}. This type of coupling is discussed in detail in the next section \ref{sec:near_wall} on near-wall interactions.
\end{enumerate}
In this work, we demonstrate examples of recent work using coupling Types 1 and 3 and discuss using MD simulations to obtain surface tension, slip-length, and contact line dynamics.
Our approach is mechanical and hydrodynamical, and we do not consider other possible approaches, for example: use of analytic method, such as solving the Ornstein--Zernike equations (for example, the Percus--Yevick closure that results in the Percus--Yevick equation) or other approximations (e.g. the hypernetted-chain equation).

\section{Results and discussion}
\label{sec:results}
In this section, we will describe developments in the use of molecular-based models and their links to continuum-scale counterparts for a range of problems, starting from relatively simple shear flows, where attention is focused on the near-wall region. Complexity is then ramped up gradually, leading up to the consideration of contact line problems involving the presence of surfactant. In each case, the type of coupling between the molecular and continuum scales is discussed, and its successes and shortcomings highlighted.

\subsection{Near-wall interactions}
\label{sec:near_wall}

It has been shown that analytical solutions for continuum flows give very good agreement with MD results even at the small scale \citep{Travis_et_al97, Todd_Daivis_book, Smith_Thesis}.
To demonstrate this, a Couette flow simulation is presented in this section, modelled by entraining a molecular liquid between two solid walls.
The molecules interact via the Mie potential, which is described by \eq{Mie_potential} using the values for Argon in Table \ref{SAFT_table} and a cutoff of $r_c = 2^{1/6}\sigma_{ArAr}$ for efficiency.
The top wall is set in translational motion and the evolution of the velocity profile towards the steady-state Couette flow limit was monitored. 
Four layers of tethered molecules were used to model each wall, with the top wall given a sliding velocity of, $U_0 = 1.0\sigma/\tau$ at the start of the simulation, corresponding to time $t = 0$ ($\tau$ is the MD time unit). 
The temperature of both walls was controlled by applying the Nos\'{e}-Hoover (NH) thermostat to the wall atoms \citep{Hoover_NoseHooverthermostat}. 
The MD simulation consists of $93, 393$ molecules, with a liquid density of $\rho = 0.4$ and solid density $1.0$.
The domain is of size $63.5$ by $46.0$ by $63.5$ in reduced LJ units split into $2$ cells in $x$ and $z$ but $512$ in $y$ to show the detailed near wall flow. 
The average density and velocity can be obtained by taking averages of the molecular values in the cells spaced over the molecular channel, in the form,
\begin{align}
\boldsymbol{u} \define \frac{1}{M_{\rm I}} \displaystyle\sum_{i=1}^N m_{\rm i} \boldsymbol{v}_{\textrm{i}} \vartheta_{\textrm{i}},
\label{U_def}
\end{align}
where $M_{\rm I}$ is the mass of molecules in the volume $I$ given by $M_{\rm I} \define \sum m_{\rm i} \vartheta_{\rm i}$, and $\vartheta_{\rm i}$ is a functional which is $1$ when the position, $r_i$, of molecule $\rm i$ is inside the volume $\rm I$ and zero otherwise. \citep{Smith_et_al2015}
This functional $\vartheta_i$ is formally the product of three boxcar functionals in each direction $\vartheta_i\define[H(x^+ - x_i) - H(x^- - x_i)][H(y^+ - y_i) - H(y^- - y_i)][H(z^+ - z_i) - H(z^- - z_i)]$ with plus and minus superscript denoting top and bottom surfaces of the volume.

\begin{figure}[H]
        \centering
(a) \hspace{3in} (b) \\
        \begin{subfigure}{0.45\textwidth}
                \includegraphics[width=\textwidth]{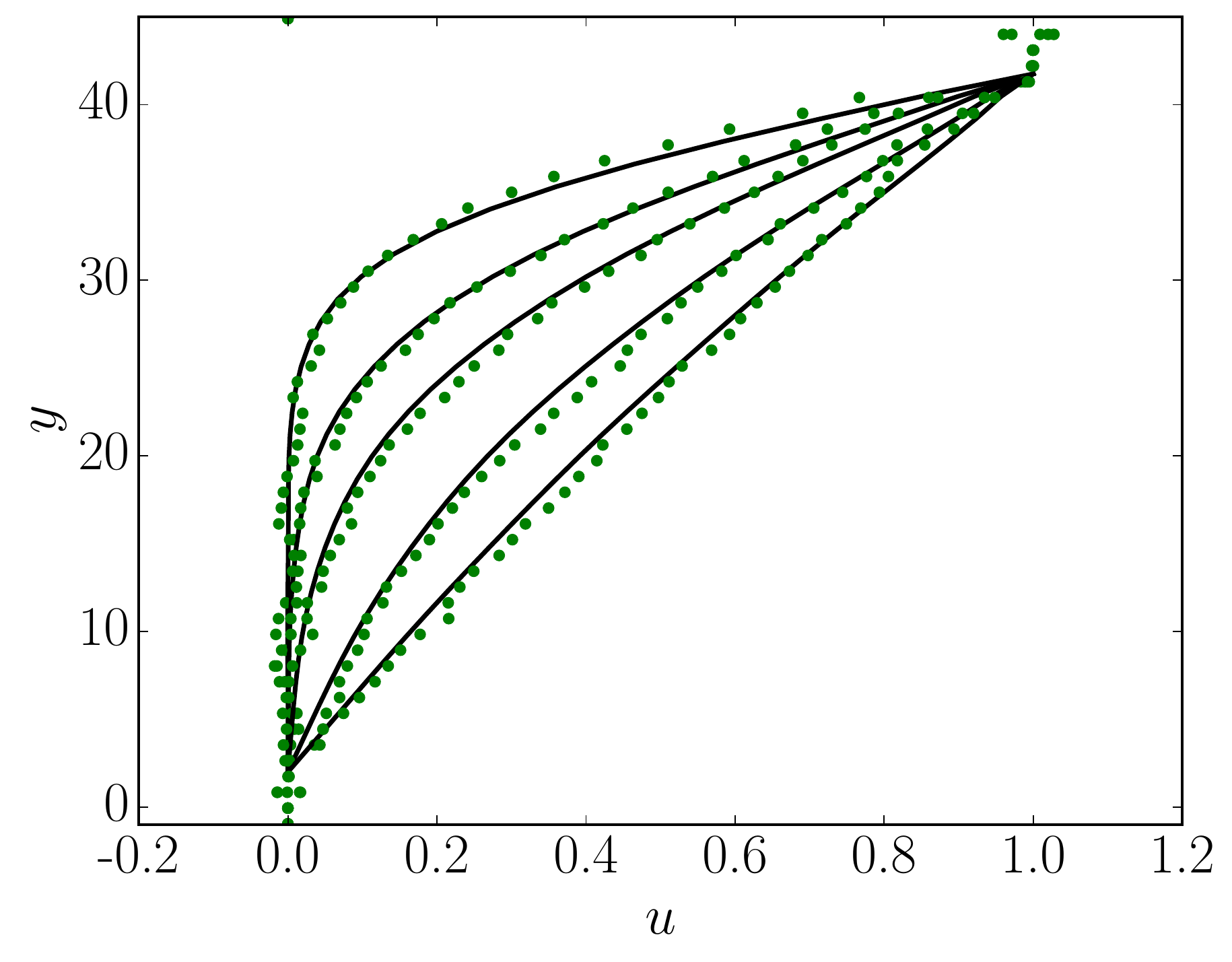}
        \end{subfigure}  \;\;\;\;\;
        \begin{subfigure}{0.45\textwidth}
                \includegraphics[width=\textwidth]{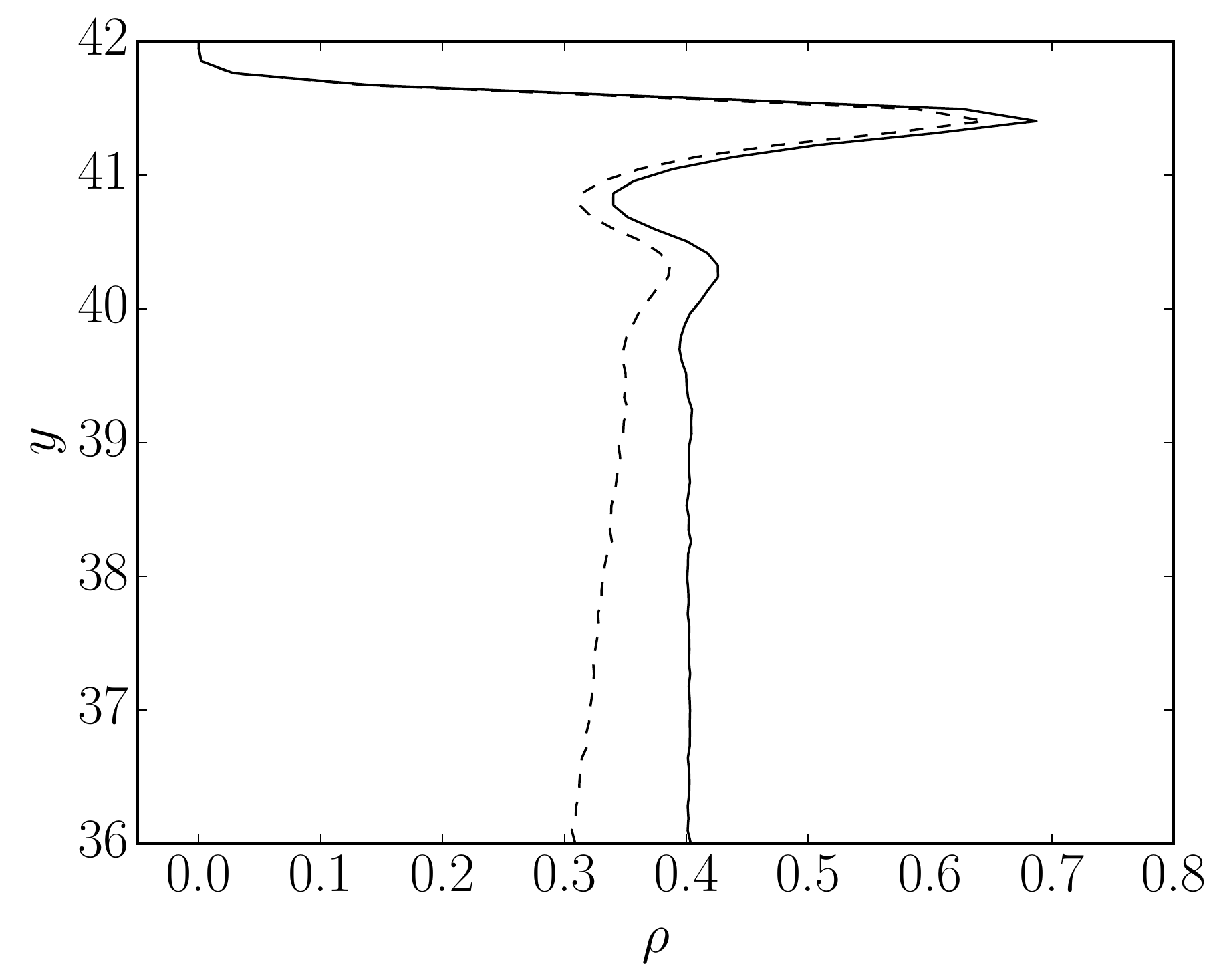}
        \end{subfigure} \;\;\; 
        \caption{Couette flow in a single-phase MD system showing excellent agreement with the continuum solution despite the small size of the system: (a) MD velocity (Green points) matched to analytical Couette flow (black line) at $t=\{ 30 ,   75 ,  130 ,  275,  530.5\}$ in reduced unit; 
(b) density (black line) and momentum  (dotted line) near the wall in the same channel where average liquid density is $\rho=0.4$. 
}
        \label{velocity_density_couette} 
\end{figure}

This velocity profile is compared with the analytical solution of the unsteady diffusion equation in Fig.~\ref{velocity_density_couette}(a), \citep{Smith_et_al} showing close agreement in both space and time.
However, some tuning of the start/end location of the analytical solution is required to get the good agreement in Fig \ref{velocity_density_couette}(a) based on the location of the walls and due to near-wall partial slip in the molecular system.
This is due to the molecular `layering/stacking' effect observed in the case of hard walls (strong tethering) and stick-slip behaviour near the walls in the molecular system, a physical phenomenon not captured by the continuum solution.
The density and momentum in the near-wall region is shown in Fig~\ref{velocity_density_couette}(b) with molecular stacking apparent for both.
This stacking effect has been observed in experiments \citep{Butt_2005} and will be important in defining near-wall dynamics of the fluid as required in defining the contact line.

Many previous studies have tried to establish a way of coarse-graining this to a simple Navier slip-length of the form of \eq{Navier_slip}. \citep{Hansen_et_al2011,Thompson1997} 
For two-phase flows, one can extract hydrodynamic boundary conditions from MD to construct a continuum model that holds in the whole space including the contact line as has been shown by Qian \textit{et al.}.\cite{Qian2003,Qian2004,Qian2006} 
In this case, the model involves a total of nine material parameters, with two of them (a mobility coefficient and a positive phenomenological parameter) matching the hydrodynamic model calculation to optimize their values.\cite{Qian2006} 
In this approach, the continuum predictions are not sensitive to these parameters within a certain range of the hydrodynamic model, where it reaches the sharp interface limit.\cite{Qian2006}
It is clear that this is possible, and indeed reasonable, for some simple fluids and surfaces, but for complex surfaces, predictions can vary by orders of magnitude \citep{Kannam_et_al2013}.

One solution to include the details of the molecular surface directly is to couple the MD and CFD descriptions (Type 3), with MD simulations near the walls and the average of this providing the CFD boundary conditions.
This type of direct coupling was originally proposed by \citet{ OConnell_Thompson} and has since received considerable attention in the literature \citep{Li_et_al, Flekkoy_et_al, Wagner_et_al, Delgado-Buscalioni_Coveney_03, Nie_et_al, Werder_et_al, Borg_et_al}. 
\begin{figure}[H]
(a) \hspace{3in} (b) \\
        \begin{subfigure}{0.44\textwidth}
                \includegraphics[width=\textwidth]{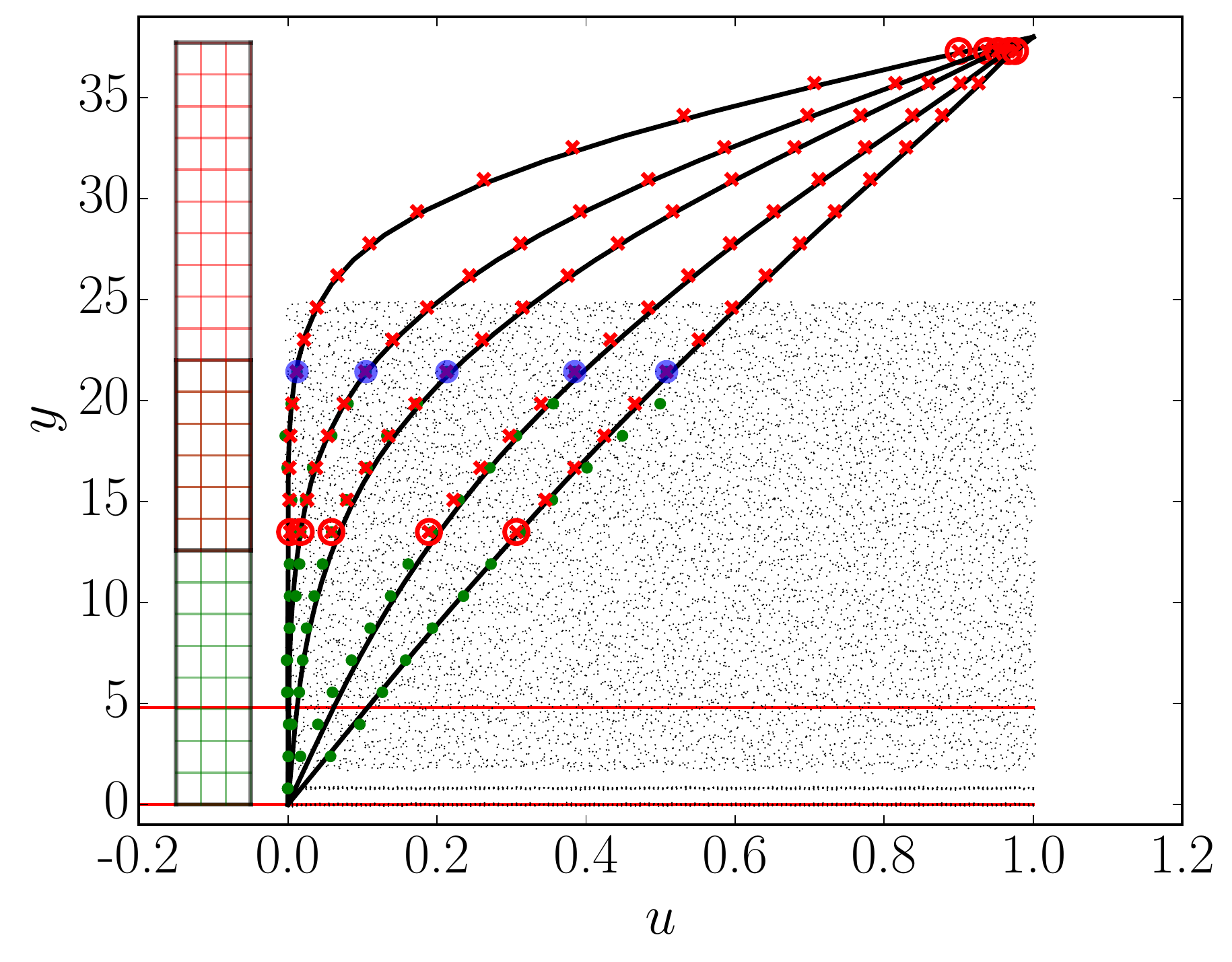}
        \end{subfigure}  
        \begin{subfigure}{0.44\textwidth}
                \includegraphics[width=\textwidth]{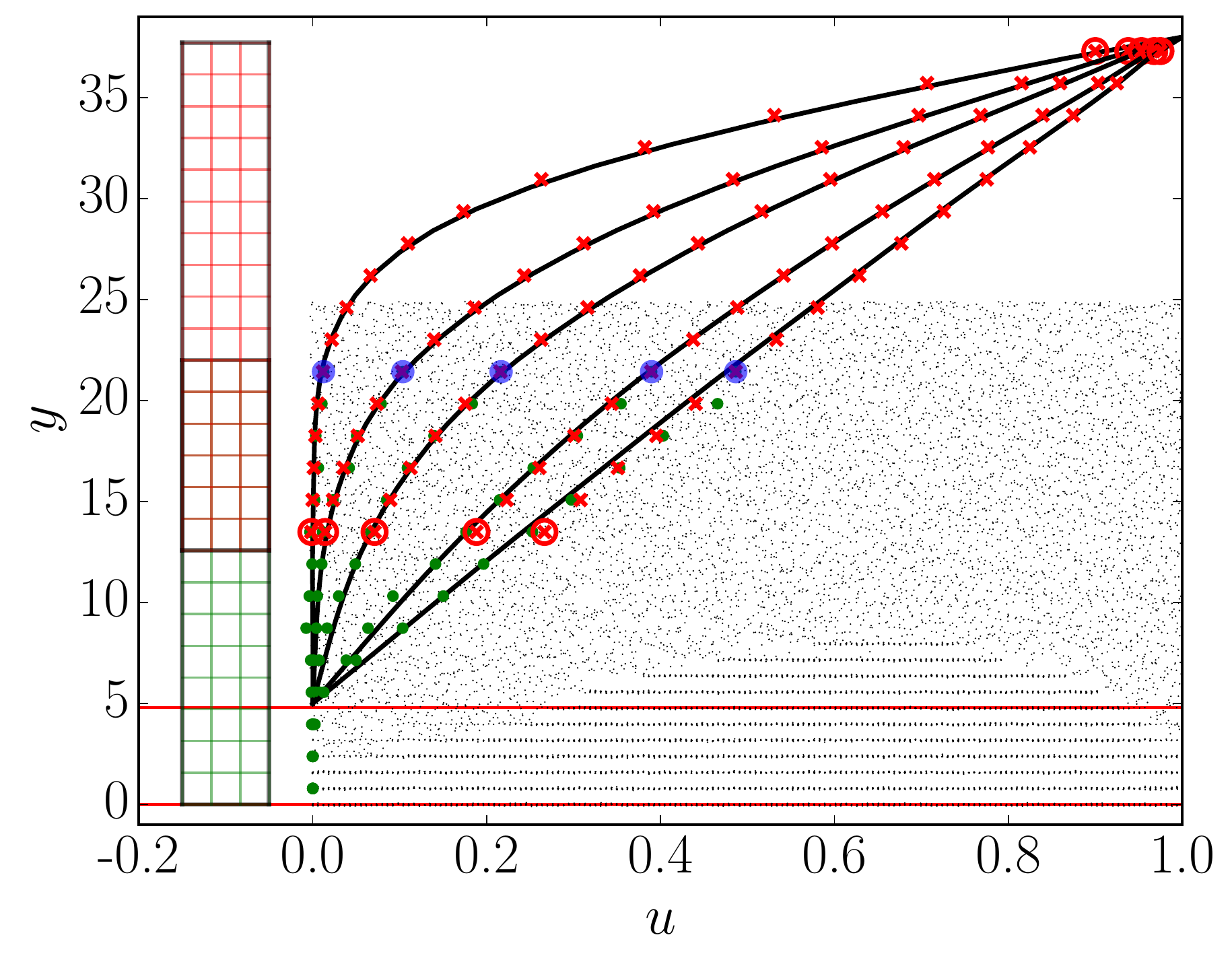}
        \end{subfigure}  
                \caption{Coupled MD--CFD model for simple LJ fluid near a wall. The different regions are MD (green $\bullet$) and CFD (red $\times$), boundary conditions averaged from the MD (red $\otimes$) and the cell constrained to agree with the CFD (blue $\bullet$) all compared to the analytical solution shown by black lines at successive times (reduced LJ units), $t=\{ 30 ,   75 ,  130 ,  275,  530.5\}$ for (a) the molecularly flat wall case and (b) wall with roughness, which shifts the effective no-slip location up to about half way across the wall, with the two horizontal red lines denoting the start of the matched analytical solution in both cases to allow comparison between (a) and (b). A snapshot of the molecules in a $10 \sigma$ wide slice in the span-wise direction are overlayed. }
                \label{CPL_velocity_couette}
\end{figure}
The bottom CFD boundary condition is obtained by averaging the velocity in the MD region as outlined in \eq{U_def} to obtain the mean velocity fields. 
The constraint applied to the MD region \citep{Smith_et_al2015} is of the form,
\begin{align}
m_{\rm i} \ddot{\boldsymbol{r}}_{\rm i} = \boldsymbol{F}_{\rm i}- \frac{m_{\rm i} \vartheta_{\rm i}}{M_{\rm I}}
\left[ \frac{d}{dt} \displaystyle\sum_{{\rm i}=1}^N m_{\rm i} \boldsymbol{v}_{\rm i} \vartheta_{\rm i} - \frac{d}{dt}\int_V \rho \boldsymbol{u} dV   \right],
\label{GLC_EOM}
\end{align}
which is Newton's law with a differential constraint, a force based on the difference in the time evolution of momentum in the overlapping molecular, $\sum m_{\rm i} \boldsymbol{v}_{\rm i} \vartheta_{\rm i}$, and continuum, $\int_V \rho \boldsymbol{u} dV$, control volumes.
This form is derived using Gauss' principle of least constraint with explicit localization using the $\vartheta_{\rm i}$ functional, which results in an extra term which was not present in previous formulations \citep{OConnell_Thompson, Nie_et_al}. 
The MD simulation consists of $122,980$ molecules for the flat wall case and $166,838$ molecules for the rough wall, with a liquid density of $\rho = 0.4$ and solid density $1.0$.
The domain is of size $130.2$ by $25.4$ by $108.0$ in reduced LJ units split into $4$ cells in $x$ each of size $32.5$, $16$ in $y$ of size $1.59$ and $4$ in $z$ of size $27.0$. 
It has the same height as the pure MD simulation shown in Fig \ref{velocity_density_couette} but split between the CFD and MD solver.
The CFD has the same number of cells $4 \times 16 \times 4$ and the same domain size, with both domains overlapping by $8$ cells. 
In Fig. \ref{CPL_velocity_couette} the grid highlights the cells used to solve the CFD and average the MD as show in red and green, respectively. The symbols at corresponding positions show the velocity in these cells.
The constraint \eq{GLC_EOM} is applied to cell $14$ of the MD domain and the top two cells are left as a buffer with the top cell thermostatted (a separate Nos\'{e} Hoover thermostat to the one applied to the bottom cell of the wall). 
This maintains the system temperature at setpoint $T_0=0.4$. 
The choice of density and temperature are based on previous simulation \citep{Smith_2015} so the viscosity is known to be $\nu = 0.7$ and this is the value set for the CFD solver so the simulations evolve together.
The simulation is run in parallel using four processes for both the CFD code, (in this case OpenFOAM \citep{openfoam} is used to solve \eq{ANDNSEqn}, although any unsteady diffusion solver would suffice), and the MD code  {\it flowmol} \citep{Smith_Thesis}. 
The parallel data exchange between these two codes is managed using the CPL library, which ensures communications are local between overlapping processors to optimise parallel scaling. \citep{CPL-library}

The agreement between the coupled model and the time-evolving Couette flow analytical solution is excellent across the two domains in Fig \ref{CPL_velocity_couette} (a).
The constraint of \eq{GLC_EOM} iterates to ensure local control of momentum in the MD simulation is exact and the agreement between MD and CFD is to machine precision, only possible by explicitly including localisation in the mathematical formulation. 
The impact of including three-dimensional wall roughness (specified using random components in Fourier space) is also shown in Fig \ref{CPL_velocity_couette} (b).
Note that as only a slice of molecules is shown, a single peak is apparent but the roughness varies in both $x$ and $z$. 
This roughness has the effect of shifting the effective location of the zero velocity point upwards into the flow.
This demonstrates the potential of this type of approach compared to parameterizing a simple Navier slip model, the actual molecular detail of wall roughness, material crystal structure, complex chemical coatings or even biological membranes can be designed and tested as part of a fluid simulation.

The Type 3 coupling described above has been applied to two-phase flows \citep{Hadjiconstantinou_thesis, Hadjiconstantinou_99} in simple channels. 
There is also work using this coupling to model droplet spreading and moving contact lines \citep{Wu_et_al14} (more on this below).
However, this required an inflow in the molecular system, which necessitated the creation and insertion of molecules.
Insertion of single atoms as well as more complicated molecules \citep{Praprotnik_2005, Borg_et_al_FADE} has been considered in the coupled simulation literature (see, for instance, Ref.~\citen{Delgado-Buscalioni_Coveney_03_USHER} and references therein).
Despite this, most MD simulations use periodic boundaries to avoid the need to specify an inflow condition \citep{Rapaport,Todd_Daivis_book,Evans_Morris}, while non-periodic boundaries require creation of information that increases the potential of non-physical artifacts being introduced. 
As a result, Type 3 coupling simulations have typically focused on flows parallel to the interface that minimize the need for insertion of molecules and modelling of a net inlet.
This form of coupled simulation is still in its infancy, and despite great progress, a clear mathematical and theoretical framework for coupling is still a way off \citep{Smith_Thesis, E_book}.

Direct coupling of this type links the dynamics of the two domains and so is limited to molecular time and length scales. 
This is a very important observation, as it means Type 3 coupling can only be used as a method of extending molecular simulations and not as a way of adding molecular detail to a continuum approach.
As a result of these limitations, in the remaining work we move away from Type 3 coupling, considering molecular modelling and development of Type 1 style couplings.
Despite these limitations, it is worth noting that the potential applications of Type 3 coupling are vast, allowing modelling of the molecular details of surfaces, nucleation of bubbles, complex chemical coatings and the flow interaction with biological membranes. 

\subsection{Surface tension effects}
\label{sec:interface}

We now consider situations in which the dynamics of a liquid--vapour interface are simulated; including the effect of surface tension. 
The transition from liquid to vapour often occurs 
over as few as three atomic layers, ~\cite{Delgado-Buscalioni2008} and it seems reasonable to assume a surface exists.
To obtain this surface, a cluster analysis is used to identify molecules in the liquid phase, with a `linked-list' built based on the criterion that molecules in a cluster are within a length $r_{\rm s} \approx 1.5 \sigma_{\rm ij}$ of each other.\cite{Stillinger1963,Theodorakis2010}
In order to define the liquid--vapour interface, the location of the edge of the liquid cluster is determined.
All molecules within a distance $r_{\rm s}$ of this surface are counted as `surface molecules'.
The surface is then defined by a function fitted to the molecular surface as points. 
For capillary wave theory, a surface is commonly defined through Fourier components \citep{Chacon2003aa}, although a similar approach using polynomial functions can remove the requirement of periodicity;
it is worth noting that the MD liquid--vapour interface is actually diffuse and any choice of surface is arbitrary, made purely as a convenient way to match to continuum concepts.

In addition to providing insight into the dynamics of the liquid--vapour surface, the surface tension can also be determined from MD simulation.
Most surface tension definitions use a thermodynamic approach, \citep{Rowlinson2002aa,Waals1979aa} based on free-energies quantities, which are not clearly defined at the molecular scale.
From a thermodynamic viewpoint, the surface tension can be interpreted as the work needed to increase the area of the dividing surface at a given curvature \citep{Ono_Kondo}. 
However, as noted in Ono and Kondo\cite{Ono_Kondo}, the thermodynamic approach to obtaining surface tension is convenient but restricted to thermodynamic equilibrium. 
Although it is possible to extend the Gibbsian approach to define curvature and contact angles \citep{Neumann_et_al}, the hydrostatic approach has the advantage that it is applicable even out of equilibrium \citep{Ono_Kondo}. 
Fluid dynamics describes the evolution of a mechanical fluid system and MD a mechanical molecular system, so the mechanical approach is used in this work.
This mechanical form of surface tension was introduced by Bakker\cite{Bakker1928} and is often referred to as the \citet{Kirkwood_Buff} method being expressed by the following formula:
\begin{equation}
\gamma = \int_{-\infty}^{\infty} \left[\Pi_{\rm N} - \Pi_{\rm T} \right] dz,
\end{equation}
where the subscript "N" indicates the perpendicular direction to the surface and
"T" is the tangential direction in the plane.
This approach, suitable for concentrations below the Critical Aggregation Concentration (CAC), requires a stress tensor $\boldsymbol{\Pi}$ to be defined in the molecular system. 
For an entire system at equilibrium, the virial approach can be used to obtain the pressure \citep{Rapaport}. 
However, the virial expression is not a valid method to get the local stress \citep{Evans_Morris, Heyes_et_al}.
Instead, the \citet{Irving_Kirkwood} (IK) approach is formally consistent with the continuum definition of stress and represents the standard microscopic formulae for computing the local expressions.
This stress is expressed here as the integral of the IK form over a volume, the so-called volume average (VA) stress \citep{Lutsko, Hardy, Cormier_et_al}:
\begin{equation}
\boldsymbol{\Pi} = \frac{1}{V} \left[ \sum_{\rm i=1}^{N} m_{\rm i}\boldsymbol{v}_{\rm i}\boldsymbol{v}_{\rm i} \vartheta_{\rm i}
 +\frac{1}{2}\sum_{\rm i,j}^N \boldsymbol{r}_{\rm ij}\boldsymbol{F}_{\rm ij}\,\int_{0}^{1} \vartheta_s ds \right].
\end{equation}
where $\boldsymbol{F}_{\rm ij}$ is the intermolecular force between molecules $\rm i$ and $\rm j$ with $\vartheta_s$ denoting a functional which is one if a fraction of the intermolecular interaction line is inside the volume and zero otherwise, defined formally in Ref.~\citen{Smith_et_al}.
The $\vartheta_s$ term is an exact analogy to $\vartheta_i$ introduced previously, with $r_{\rm i} \to r_s$; the location of a point on the line of interaction between molecules $r_s = r_{\rm i} + sr_{\rm ij}$  is used instead of particle position. The dummy variable $s$ is integrated between $0$ and $1$, corresponding to tracing the line between molecule $\rm i$ and $\rm j$ to obtain the fraction of the intermolecular interaction inside the volume. 
Considering stress over an integrated control volume removes some of the ambiguity in its definition (see \eg \citet{Zhou, Admal_Tadmor})  by allowing direct comparison to the control volume form of the equations (\ref{BofMEqn2}) \citep{Smith_et_al}.
The non-uniqueness of stress is important when the actual value of surface tension depends on stress in the vicinity of the interface.
Recent work has shown that the distributions of measured stresses in volumes $V$ smaller than $V^{1/3} = 3$ are non-Gaussian, suggesting that simply taking mean values of stress may not be sufficient\citep{Smith_et_al2017}.
This is a concern as the stress within only a few intermolecular diameters of the surface can be shown to almost entirely determine the surface tension \citep{Delgado-Buscalioni2008}.
In addition, three-body interactions or even more complex multi-body potentials are needed even for simple LJ, as shown by a recent study \citep{Ghoufi_Patrice_Tildesley}.
It is clear that care is required in defining both a vapour--liquid surface and its tension.

	\begin{figure}
        \centering
(a) \hspace{3in} (b) \\
        \begin{subfigure}{0.45\textwidth}
                \includegraphics[width=\textwidth]{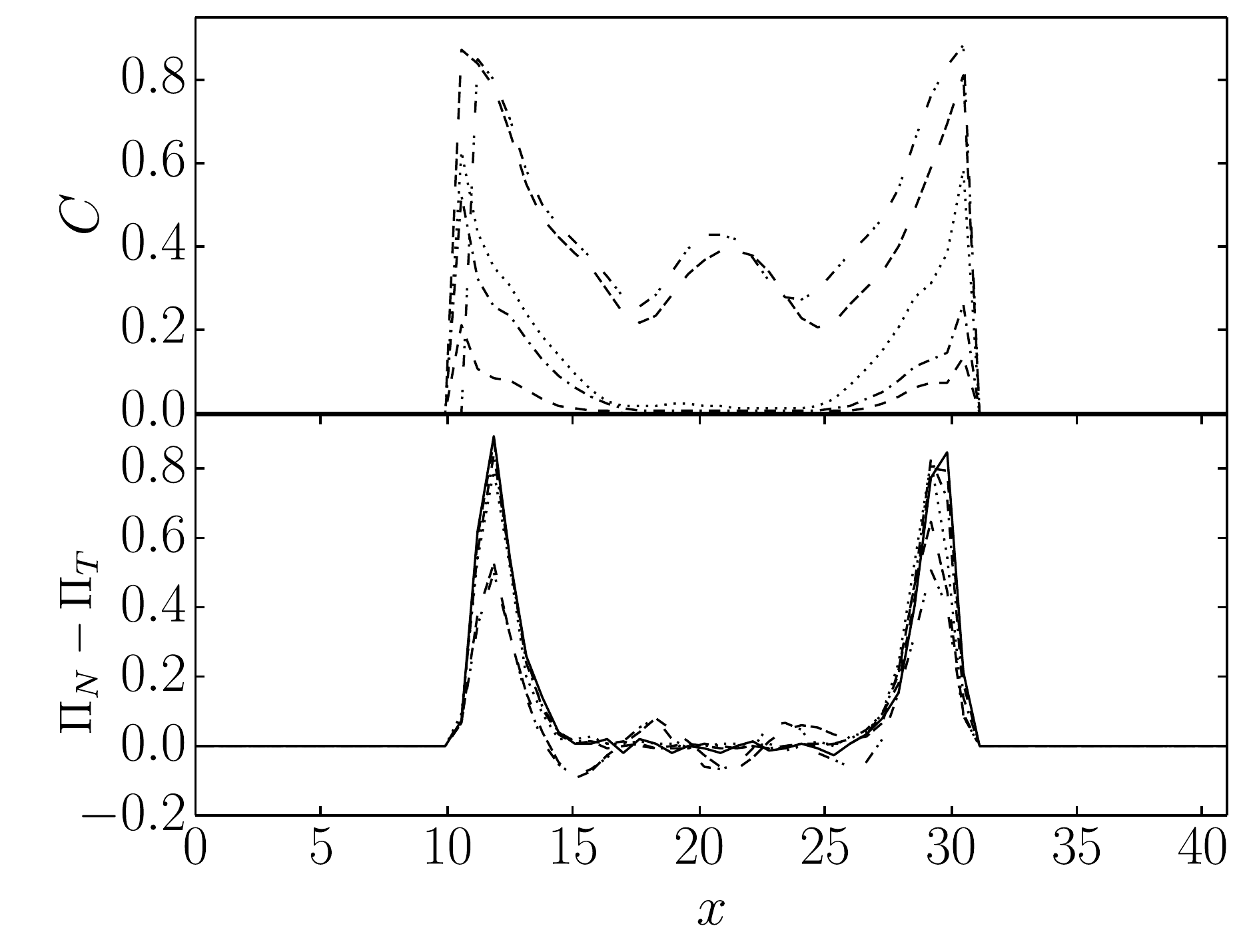}
        \end{subfigure}  \;\;\;\;\;
        \begin{subfigure}{0.45\textwidth}
                \includegraphics[width=\textwidth]{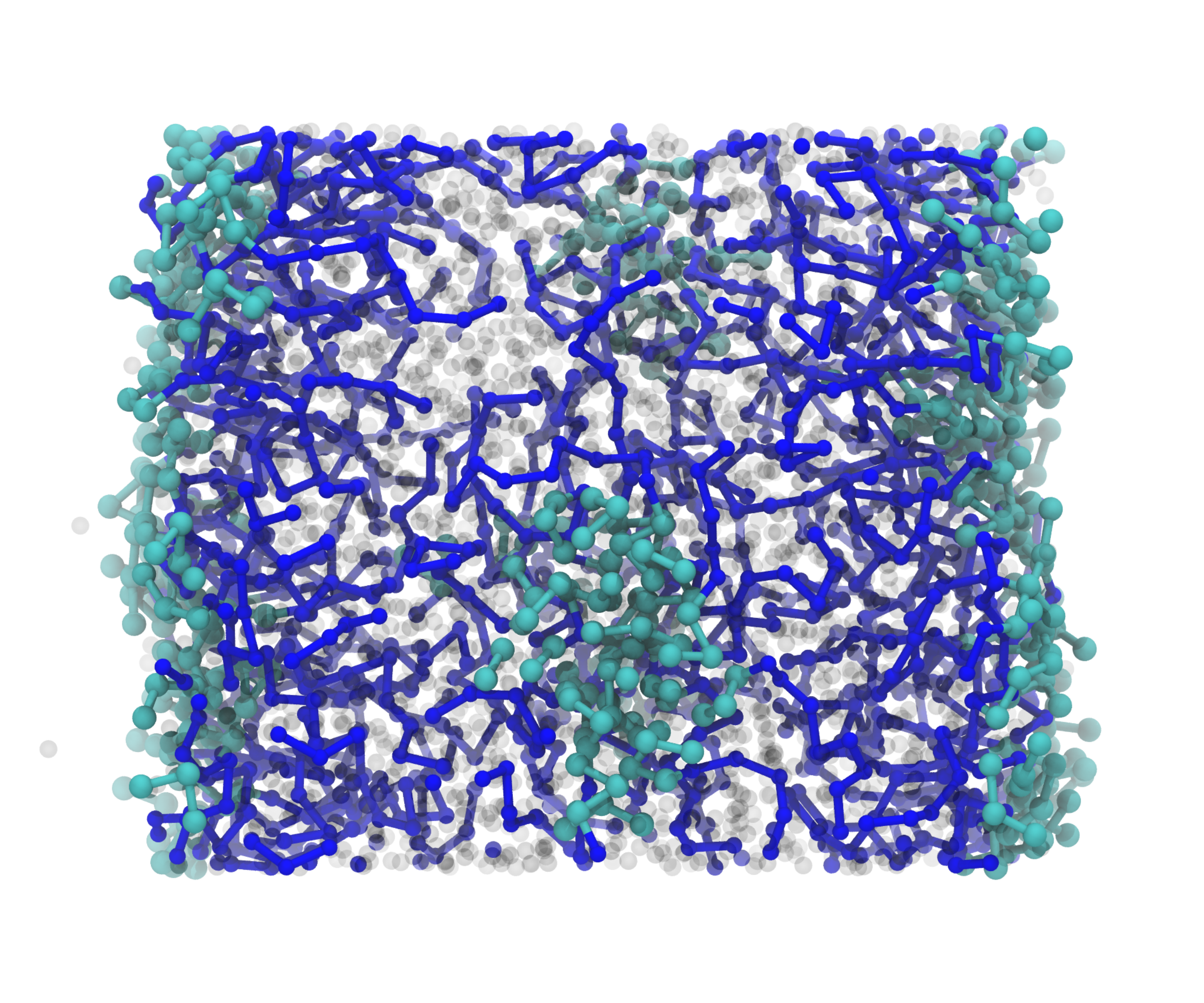}
        \end{subfigure} \;\;\; 
(c) \hspace{3in} (d) \\
\begin{subfigure}{0.45\textwidth} 
                \includegraphics[width=\textwidth]{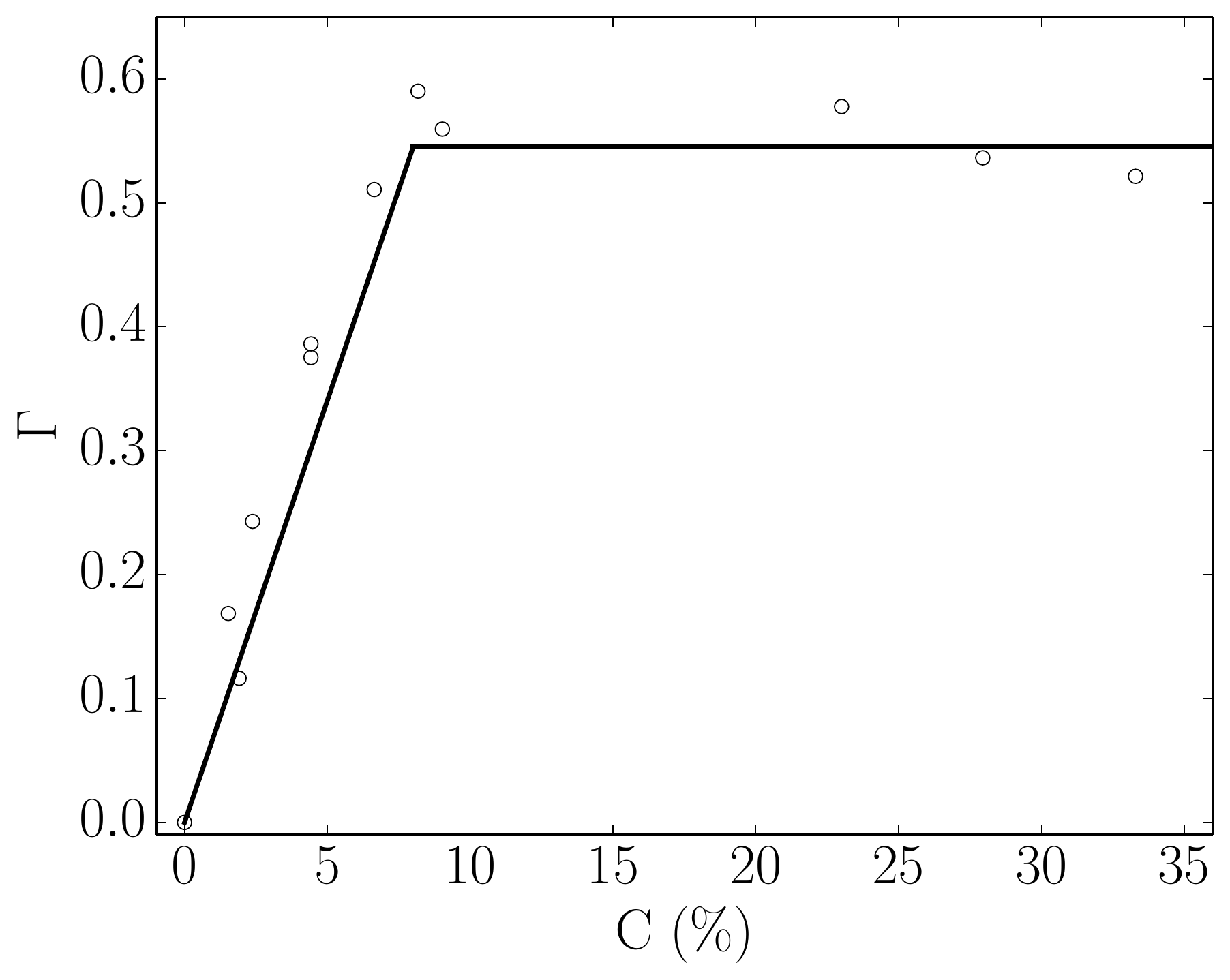}
        \end{subfigure} 
        \begin{subfigure}{0.48\textwidth}
                \includegraphics[width=\textwidth]{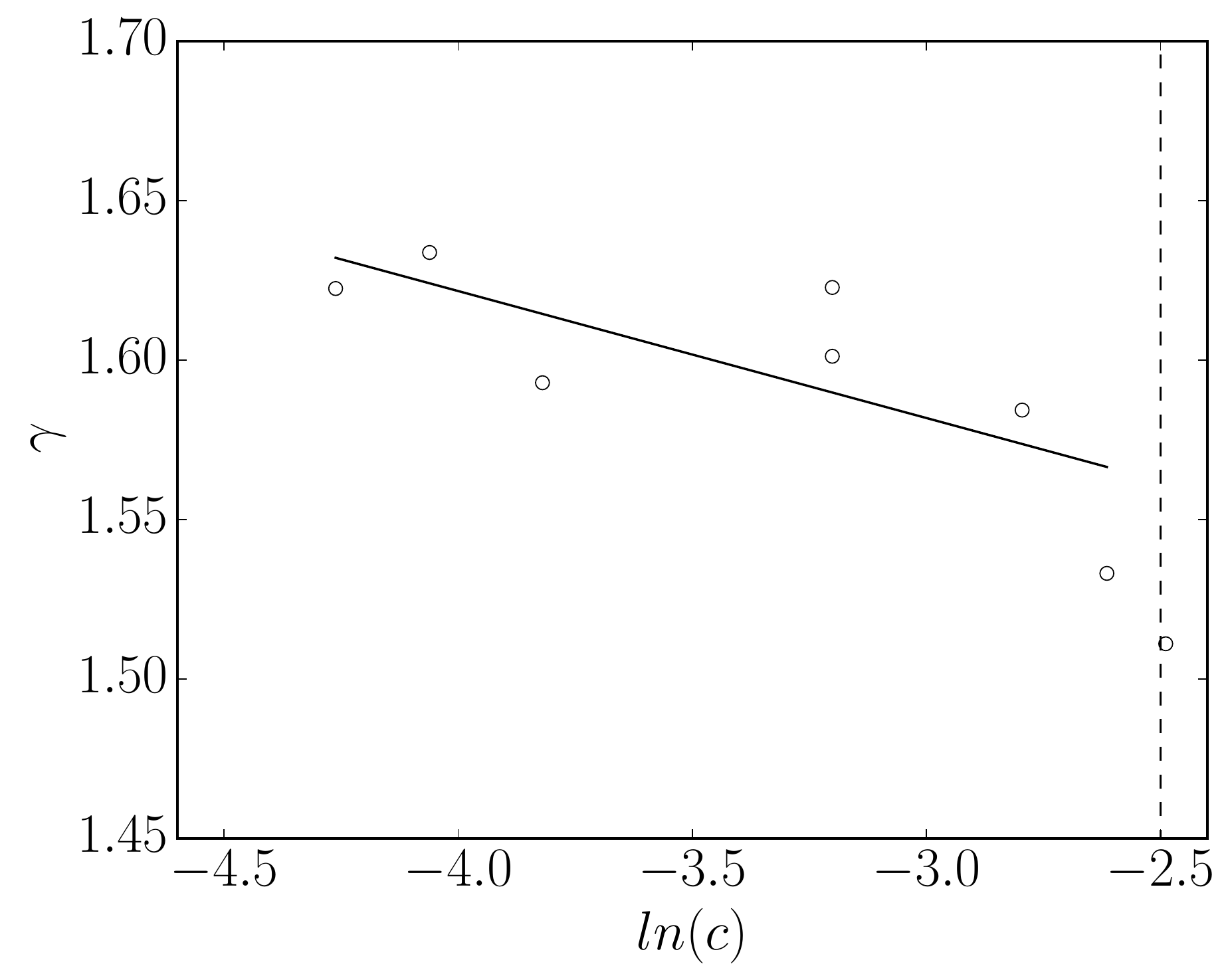}
        \end{subfigure} 
                        \caption{Molecular dynamics simulations of interfaces with surfactants: (a) (top) concentration of surfactant, $C = \rho_{poly}/\rho_{Tot}$ where $\rho_{Tot} =  \rho_{poly} + \rho_{solv}$ with $\rho_{poly}$ being the density of surfactants and $\rho_{solv}$ being 
the density of the solvent (water), and (bottom) contribution to surface tension integral $\gamma$ from differences in pressure as a function of domain position; (b) two-phase water (transparent) and surfactant (blue hydrophilic tails and teal hydrophobic heads) in a simulation used to calculate the surface tension;
(c) surface excess concentration $\Gamma \define C_{Surface} - C_{Bulk}$ as a function of concentration, which suggests that the critical aggregation concentration (CAC) is around $8\%$; (d) surface tension $\gamma$ in reduced units as a function of concentration; here, the dotted line is at the CAC of $8\%$ with points above this value omitted from the figure. The surface tension in the limit of zero concentration, $\gamma \approx 1.64$ corresponds to the standard temperature and pressure surface tension of water, $71mN/m$, as the SAFT model is designed to give this \cite{Lobanova2015,Theodorakis2015,Theodorakis2015b}.
}
                \label{vmd_Surface_concentration}

\end{figure}

Despite the aforementioned challenges, molecular simulation is uniquely placed to provide \textit{a priori} estimates of surface tension for complex multi-component systems such as surfactant-laden fluids.
Liquid--vapour interfaces can be designed based on the required molecular structures, downloaded from online databases \citep{PDB} which are tuned from quantum potentials or experiments, and verified against X-ray scattering data \citep{Hansen_Mcdonald, Yamell_et_al}.
The bulk behaviour of systems combining different numbers of these molecules allows the exploration of the effect of their concentration on the surface tension.
To demonstrate this, the surface tension contribution as a function of spatial location is shown for a range of surfactants concentration in Fig \ref{vmd_Surface_concentration} (a).
The modeled system is two-phase water with varying concentrations of a SAFT based model for an organic poly-alkyl-ether molecule, ($CM-CM-CM-EO-EO-EO-EO-EO-EO-EO-EO$ with $-$ denoting harmonic bonds, see table \ref{SAFT_table}). 
The concentration of surfactant at the surface can be seen to increase for greater concentrations in Fig \ref{vmd_Surface_concentration} (a) (top) while the contribution to surface tension at the surface, the difference in normal and tangential stress $\Pi_N - \Pi_T$ is reduced by the increased surfactant concentration (bottom). 
The results of increasing surfactant concentration is also shown in Fig. \ref{vmd_Surface_concentration} $(c)$ for the surface excess against concentration. 
It is clear that beyond the critical aggregation concentration (CAC), approximately $8\%$ in this case,  no more surfactants can be accommodated at the interface.
The measured surface tension is shown as a function of this concentration in  Fig. \ref{vmd_Surface_concentration} $(d)$. 
For concentrations above the CAC, surface tension continues to drop, which highlights a feature of the \citet{Kirkwood_Buff} formulation as it includes the integral over the whole domain, including the increasingly surfactant-laden bulk.
This bulk is increasingly inhomogenous as micelles form, creating new surfactant water interfaces inside the liquid region, and the $\Pi_N - \Pi_T$ term will no longer be zero on average inside the liquid.


%
%
%
%


\subsection{Static contact angles}
\label{sec:wall_interaction}

Having considered details of both the wall--fluid interaction and the liquid--vapour interface, we move on to consider the point where both meet, the contact line. 
In a recent experimental study, \citet{Nelson_et_al_2011} explored the impact of electrowetting and wall sliding speed on dynamic contact angle.
Despite the small system size in the case of molecular modelling, good agreement between molecular-level simulation and experiment has been observed.\citep{Cheng_et_al2016}
The static contact line is analyzed first, to parameterize this behaviour before the added complexity of moving the contact line is considered.


In a molecular model, wall-fluid interactions can be varied by changing the interaction, $\epsilon_{\rm wall}$, between the wall and the fluid molecules.
There is a large number of factors that determines surface--fluid interplay in experiments, including surface roughness and material, complex chemical coatings as well as electrowetting.  
We consider the flows observed in experiments \citep{Nelson_et_al_2011} in which a liquid bridge is sheared between a gold surface on the top and a sliding Electro-Wetting on Dielectric (EWOD) surface on the bottom; here, electrowetting effects are used as a means of effectively-varying the wall--fluid potential, which, in turn, alter the wall wetting properties \cite{Theodorakis2015b}.

\begin{figure}
        \centering
(a) \hspace{2.8in} (b) \\
        \begin{subfigure}[t]{0.48\textwidth}
		\includegraphics[width=\textwidth]{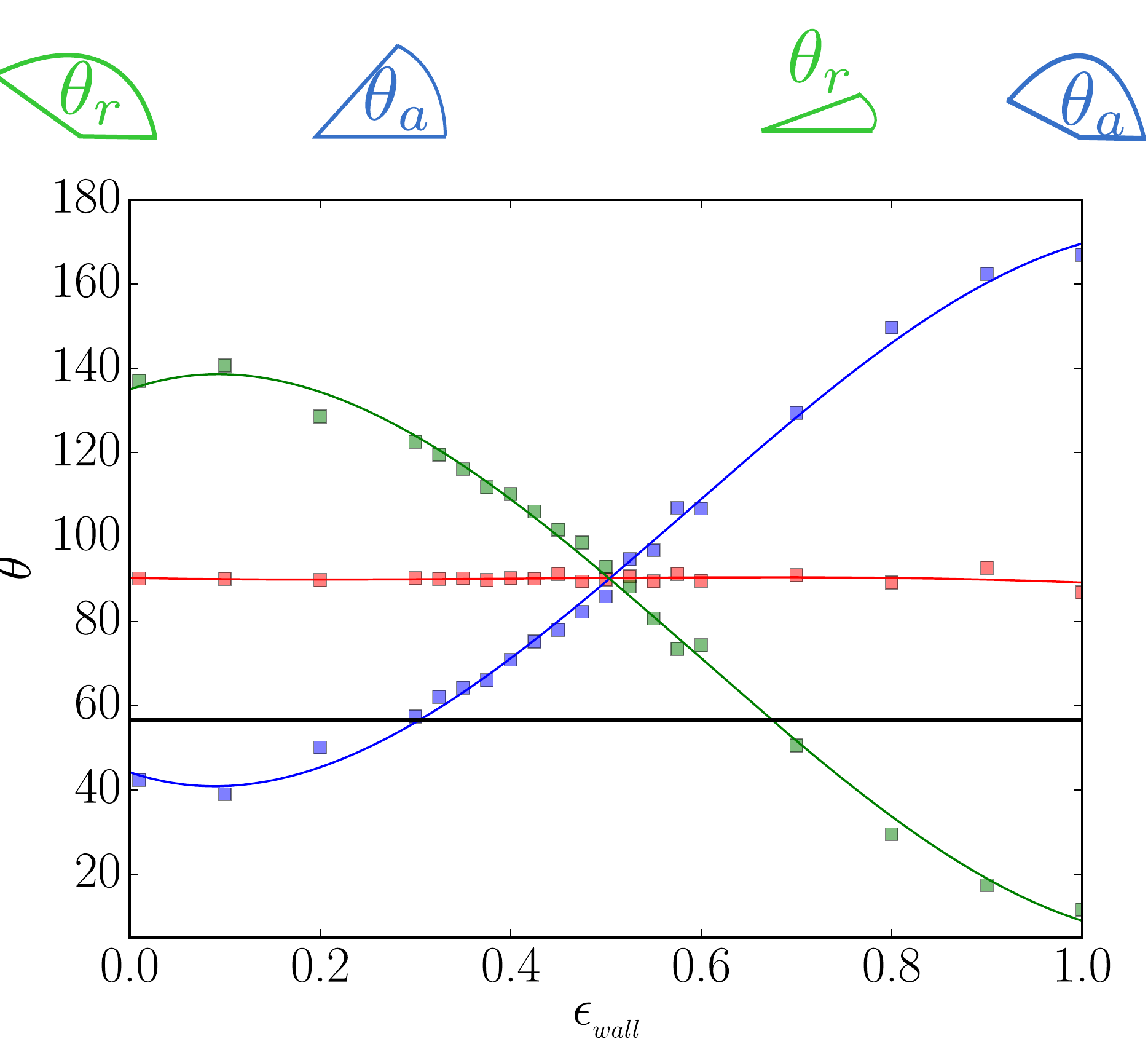}
         \end{subfigure} 
                 \begin{subfigure}[t]{0.48\textwidth}
                 \includegraphics[width=\textwidth]{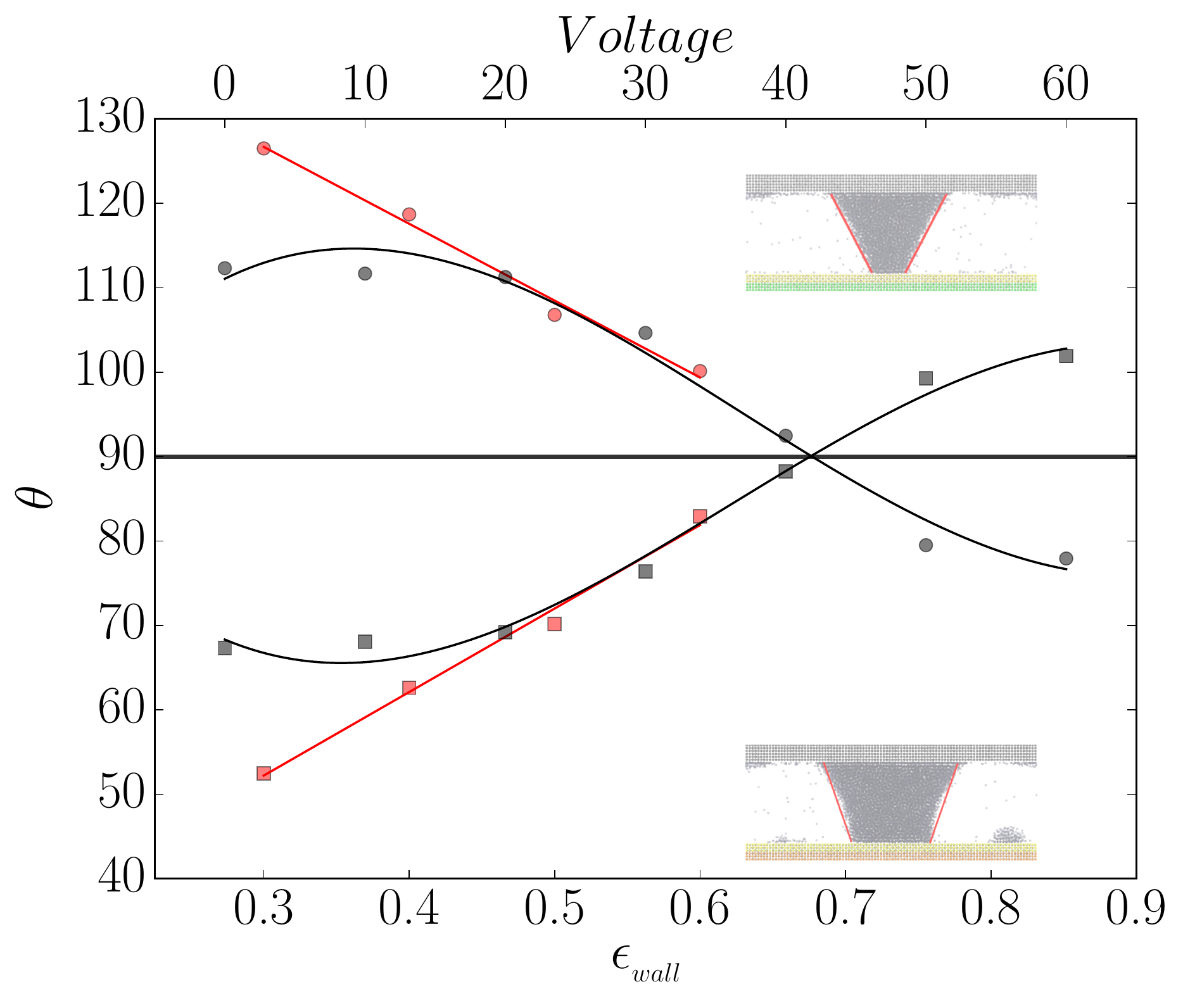}
         \end{subfigure} 
       \caption{Static contact angles as a function of wall wetting value $\epsilon_{\textrm{wall}}$, (a) the same $\epsilon_{\textrm{wall}}$ for top and bottom walls obtained from a cubic line fitted to the surface  with advancing angle (blue), receding angle (green) and slant angle from a linear fit (red). The three cases are shown above the figure with the fitted lines used to determine angles shown (slightly unusual choice of measuring anti-clockwise for the bridge, with angle determined from the right at the bottom, left from the top) and the measured angles, while the horizontal black line indicates the contact for water--gold interaction angle \citep{Garcia_et_al}. (b) Using $\epsilon_{\textrm{wall}}^T=0.7$ for the top wall (based on gold--water angle), the slant angle from the linear fit for varying bottom wall interaction $\epsilon_{\textrm{wall}}^B$ (red) measured with the same convention (anti-clockwise) as the advancing and receding angles, is compared to experimental data with second x-axis for Voltage (black) \citep{Nelson_private_communication, Nelson_et_al_2011}.Adapted with permission from Ref. 31. Copyright 2016 Royal Society of Chemistry.}
         \label{fig:wetting_study} 
\end{figure}

First, the results of a parametric study of wetting using MD simulations with a symmetric wall for a range of different wall interactions are presented in Fig \ref{fig:wetting_study}(a); here, the SAFT form of water from table \ref{SAFT_table} is used for the liquid and vapour molecules.
It is worth noting that the extreme values are not strictly correct due to the use of a cubic polynomial to define the liquid--vapour surface.
The angles are, therefore, over-predicted or under-predicted for the hydrophillic and hydrophobic case, respectively. 
These extreme values are not relevant, and so this discrepancy is not important for the results presented here.
From the work of \citet{Garcia_et_al}, the contact angle of water on a gold substrate is $56.6^\circ$ at room temperature (shown as a horizontal line on Fig \ref{fig:wetting_study}(a)).
In order to reproduce this gold--water contact angle, a wall interaction of $\epsilon_{\rm wall}^T = 0.7$ is chosen and the top wall set to this value, while the bottom wall interaction is varied separately to match the static contact angle from \citet{Nelson_et_al_2011}.
These static contact angles in the molecular systems are parameterized and the overall trends compared to the experimental results for different electrowetting cases in Fig \ref{fig:wetting_study}(b).
From this, four values, $\epsilon_{\rm wall}^B = \{ 0.38, 0.46, 0.52, 0.69\}$ are chosen to match $10V$ to $50V$ from the electrowetting study.
We note that, in principle, voltage controls the macroscopic contact angle as expressed by the Young--Lippmann equation (YLE), while the interaction energy parameter is related to the microscopic contact angle. 
Although, this may be misleading when matching dynamic contact angles, \citet{Liu2012} present results which ``provide strong evidence that the YLE remains valid down to nanometer scales''. 
In order to obtain a simple model of contact line motion in this work, we tune the interaction strength, as an analogy to voltage in the YLE, to manipulate the microscopic contact angle.  
The range of wall interactions covers the same angles observed in the experiments of \citet{Nelson_et_al_2011} except large voltage values which exhibit unstable behaviours.
Having matched the static starting angles, a parametric study of wall sliding speeds is performed in the next section.

\subsection{Dynamic contact angles}
\label{sec:dynamic_contact}

Molecular dynamic studies of the moving contact angle using sheared liquid bridges have shown great promise in the literature, matching Cox--Voinov \citep{Cox_1986a} and molecular Kinetic Theory, \citep{Blake_2006} but have not been directly compared to experimental results. \citep{Thompson_et_al89, Thompson_et_al93, Gentner_et_al_2003, Qian_et_al_2003, Ren_E_2007}
In this section, we are interested in the similarities and differences of these molecular simulation to results from similar experimental geometries.
The setup in this section is the same as in the previous section, represented by Fig. \ref{fig:wetting_study} (b), except that now the bottom wall is allowed to slide.
\begin{figure}
        \centering
(a) \\
        \begin{subfigure}[t]{0.65\textwidth}
	  \includegraphics[width=\textwidth]{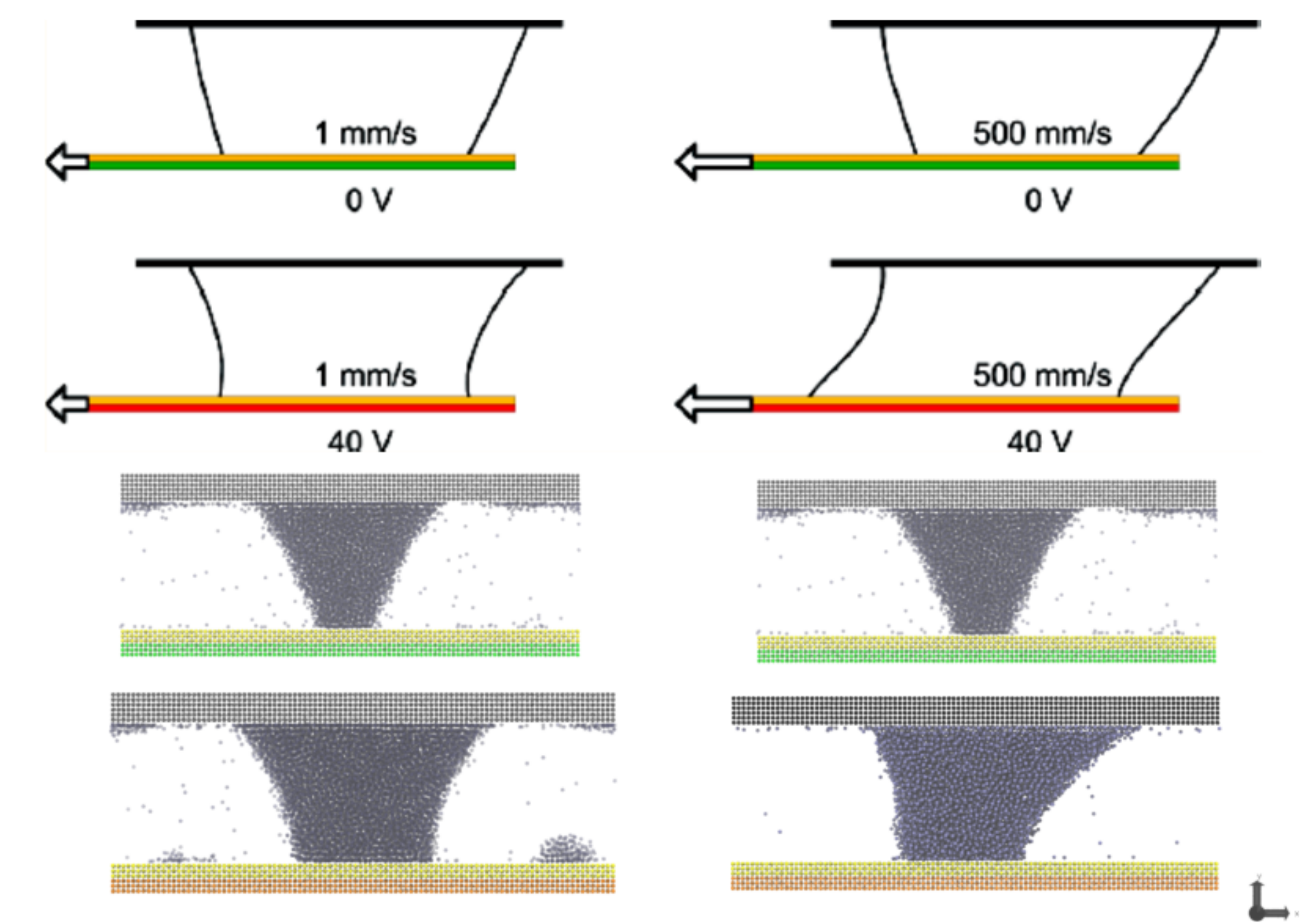}
        \end{subfigure}  \\
(b) \hspace{3in} (c) \\
        \begin{subfigure}[t]{0.65\textwidth}
	    \includegraphics[width=\textwidth]{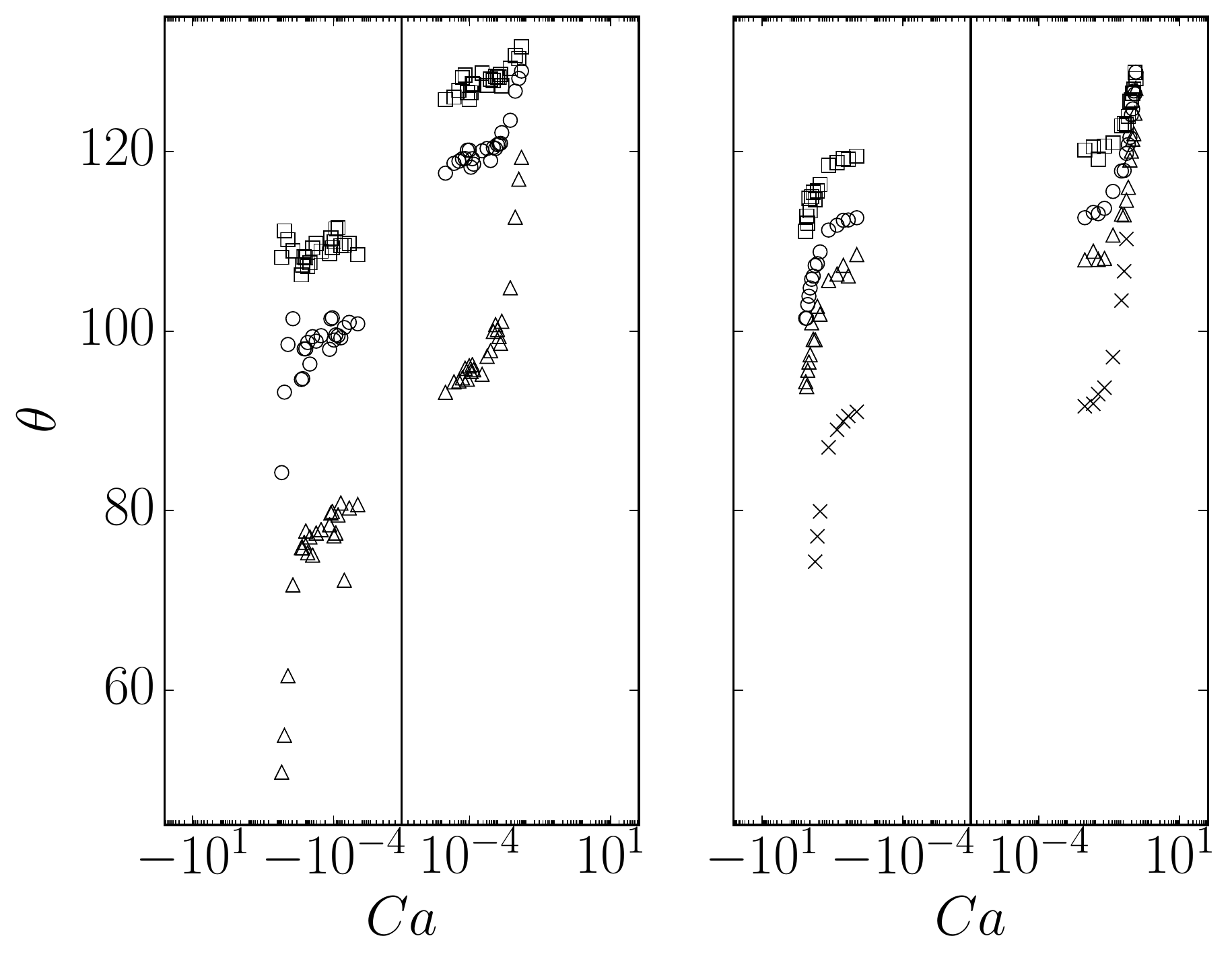}
         \end{subfigure} 
    \caption{(a) Qualitative comparison of contact angle sliding for different voltage and Capillary number as indicated for analytical and molecular modelling results (using the same surface tension $\gamma=71.97\times 10^{-3}N/m$ and viscosity $\mu = 1.002\times10^{-3} N/(s m^2)$ for both MD and experiments).  Result for the contact angle $\theta$ vs. Capillary number with (b) from experiments of \citet{Nelson_et_al_2011} with EWOD voltage $0V$ ($\square$), $20V$ ($\circ$) and $40V$ ($\triangle$) and (c) wall interaction $\epsilon_{\textrm{wall}}$ of 0.38 ($\square$), 0.46 ($\circ$), 0.52 ($\triangle$) and 0.69 ($\times$).Adapted from Ref. 114. Copyright 2011 American Chemical Society.}
    \label{Nelson_et_al_2011_vs_MD}
\end{figure}
The static contact angles were matched in the previous section for a range of electrowetting strengths.
The dynamic behaviour of these matched static contact angles are compared between MD and experiments in Fig \ref{Nelson_et_al_2011_vs_MD} (b).
The experimental result for the fastest sliding speed from \citet{Nelson_private_communication} overlaps with the slowest MD result.
The hysteresis seen in the experimental results is significant, with almost $20^\circ$  in the slowest sliding case, while almost no hysteresis is observed in the MD simulation.
This difference in the contact angle hysteresis may be attributed 
to the small system size in the case of the 
simulation \cite{Extrand1995,Whyman2008,Park2015,Eral2013}, which is predicted to be more 
pronounced for smaller systems.\cite{Whyman2008}
At sufficiently large speeds, there is a minor hysteresis in the advancing and receding angles, as shown in Fig.~\ref{Nelson_et_al_2011_vs_MD} (a).

The molecular liquid bridge does not remain stationary and travels along the domain, which most likely reduces hysteresis as the contact line is not pinned. 
The movement of the liquid bridge does not occur with counter-sliding walls, and appears to result from the asymmetry of a stationary top wall and moving bottom wall.
Furthermore, quantitative agreement is not observed between MD and experiments. 
Possible reasons for this discrepancy include: $i)$ the simplicity of the MD model for water (SAFT here, SPC or TIP4-P may be better), $ii)$ inadequacies of using interaction strength to model electrowetting, $iii)$ the perfect atomic lattice used for the walls, and iv) system size effects.
Considering point $i)$, the force-field based on the SAFT-$\gamma$ equation of state does not have electrostatic terms and has not been fitted to hydrodynamic properties such as viscosity, but is instead a result of fitting the equation of state to experimental thermodynamic data. 
Hence, a possible refinement of the fitting parameters could also lead to a more accurate model.
For point $ii)$, using an actual model of the electrowetting force-field instead of simply tuning the fluid--wall interaction, as in \citet{Zhao_Quanzi}, or a more sophisticated force applied to all molecules. \citep{C4NR06759B} 
For $iii)$, the lattice is very different from the chemical and physical roughness observed in the experiments which would pin the contact line to a particular location. In fact, even the actual tethering of the wall molecules can be an important factor in the dynamics of a contact line.
In our case, the tethering of molecules leads to stiff substrates, but one can vary the strength of the tethering interactions creating soft and hard substrate areas as appropriate.
This property has been recently used in simulation `experiments' to guide droplets along stiffness gradients on solid substrates; a phenomenon known as {\it durotaxis}. \cite{Chang2015,Theodorakis2017} 
By varying the stiffness of the substrate, and up to a threshold value for the stiffness, the contact angle of a droplet can change significantly (\eg by as much as 20$^{\circ}$).\cite{Theodorakis2017}

However, the limited system size may be the main source of error, as it is insufficient to accommodate the scales of motion necessary to provide a faithful representation of the dominant physics.\cite{Whyman2008,Extrand1995}
Eral \textit{et al.}\cite{Eral2013} notes that MD simulations have tried to model hysteresis but size and length scales are too limited by computer costs. As will be shown in the next section, MD system sizes must exceed a certain size before agreement with continuum-scale models is observed.\cite{Theodorakis2015b} This observation is supported by the experiments of Park \textit{et al.}\cite{Park2015} where below 5 μm the spreading behavior of the contact line is shown to be very different, with nano-scale wall roughness being an important factor.

Despite the differences between experiments and MD, the observed shapes and dynamic trends of the liquid bridge are still seen to be broadly similar to experimental observations in Fig \ref{Nelson_et_al_2011_vs_MD} (b).
In particular, inspection of this figure reveals that the advancing contact angles converge 
at high speeds, and the receding ones diverge, 
with a larger degree of wetting effectively shifting the curve down in both the numerical and experimental results.
The sheared liquid bridge has the advantage that it is a steady state, allowing us to collect detailed statistics on the behaviour of the contact angle.
As a result, the molecular dynamics simulation of a sheared bridge can provide insight into the fluctuations due to molecular motions \citep{Smith2016, Smith_et_al2017}.
This provides an opportunity to develop a reduced model for these molecular fluctuations in the spirit of Type 1 coupling.
As shown in recent work  \citep{Smith2016}, these fluctuations are largely Gaussian and the autocorrelation is well described by exponential decay.
As a result, a Langevin equation can be used to reproduce the molecular behaviour at the contact line by tuning it with the MD mean, standard deviation, and autocorrelation,
\begin{align}
\theta^{t+1} = \theta^t -  \frac{k \Delta t}{\Gamma} \left[\theta^t - \langle\theta\rangle \right] + \xi \frac{\sqrt{C \Delta t}}{\Gamma}. 
\label{Langevin_equations_numerical}
\end{align}
Using \eq{Langevin_equations_numerical}, molecular fluctuations can be incorporated into a CFD model as a form of Type 1 coupling.
This was demonstrated in \citet{Smith2016} with evolution of the mean velocity, $\langle\theta\rangle$, governed by a form of Tanner's law and the Langevin equation used to advance the continuum angle but including molecular noise.

\begin{figure}
(a) \\
\includegraphics[width=0.7\textwidth]{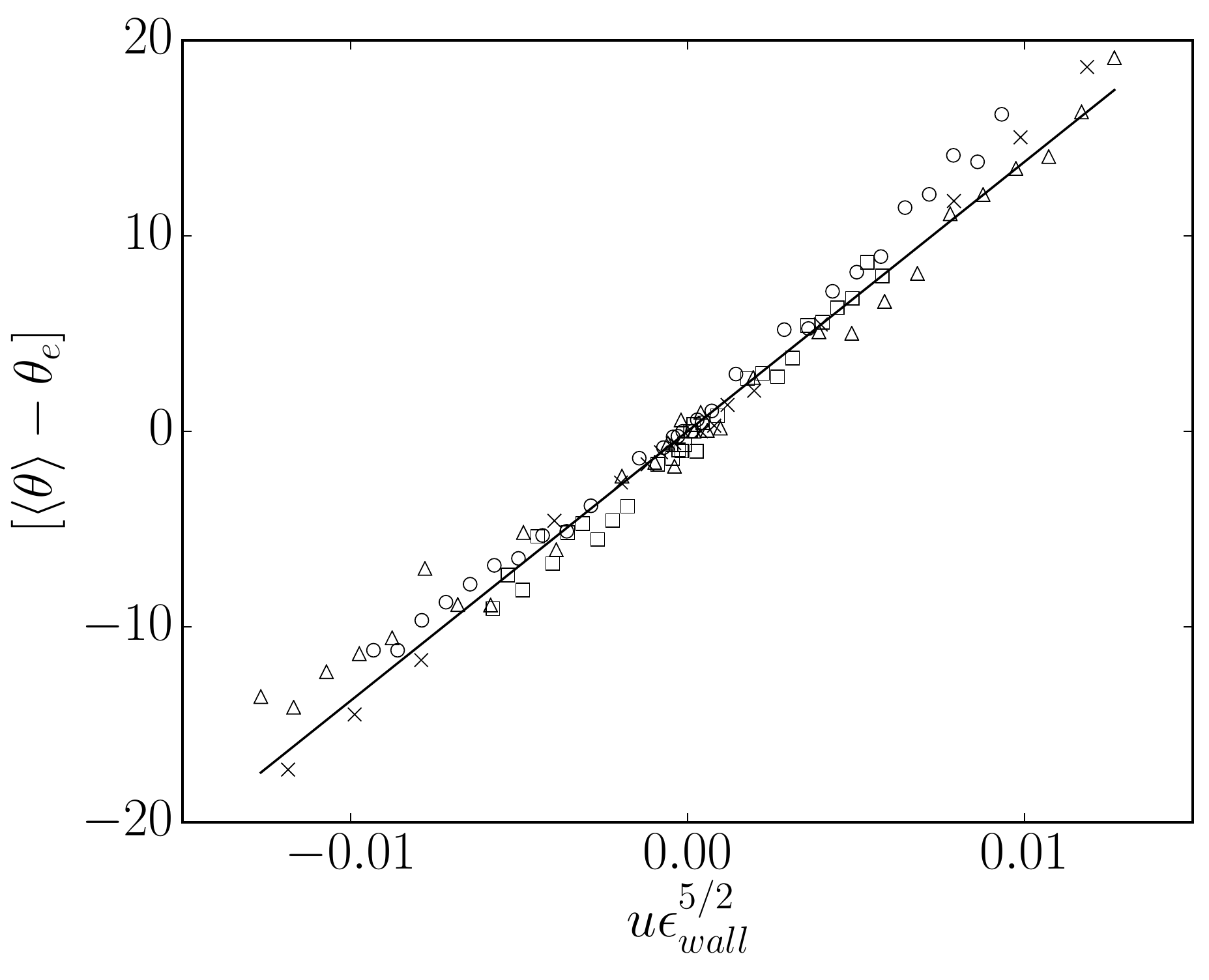}
\\ (b) \\
\includegraphics[width=0.7\textwidth]{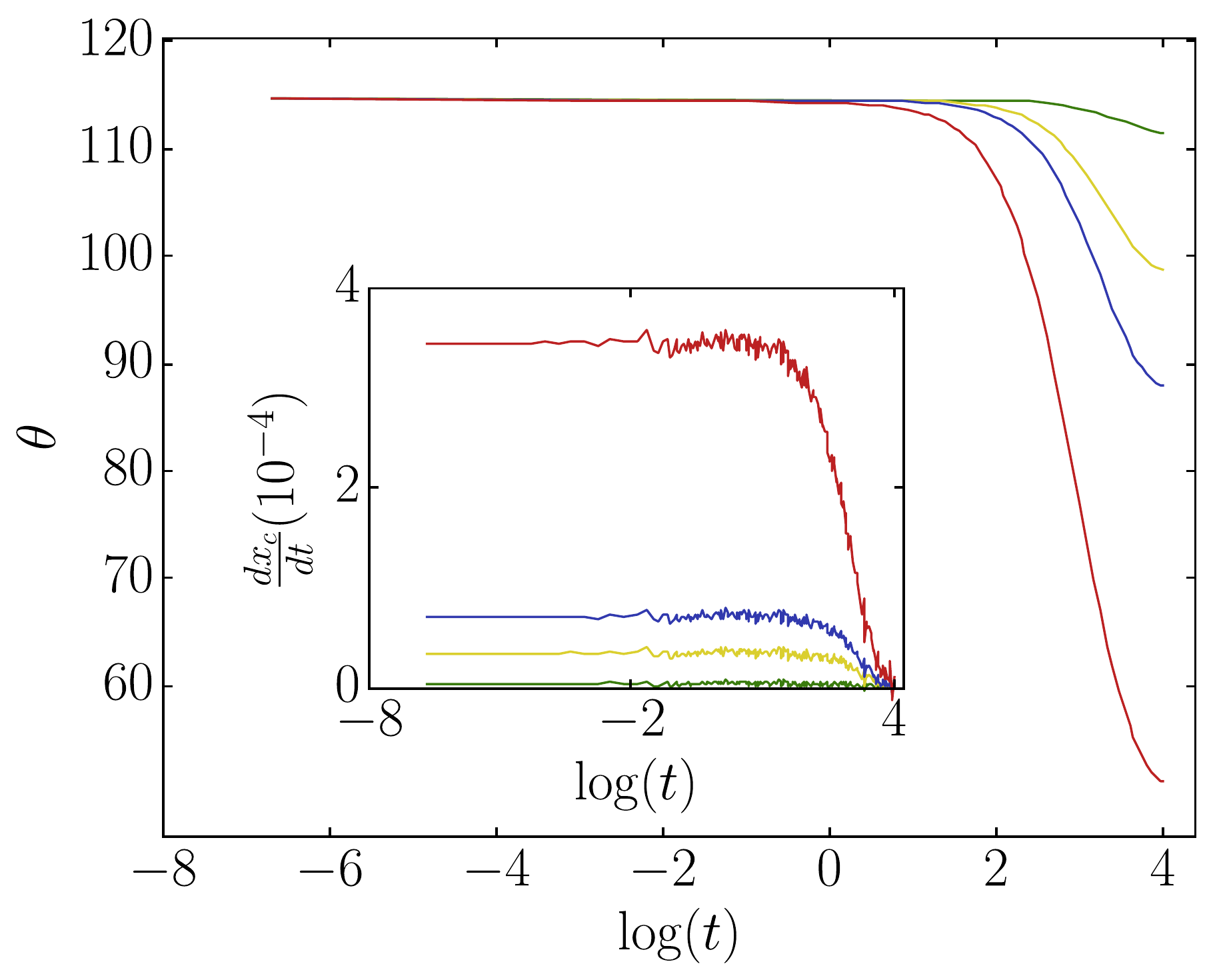}
\caption{(a) Wall sliding speeds for four wetting numbers collapsed onto a single line by scaling using $\epsilon^{5/2}_{\textrm{wall}}$. (b) CFD solver using the thin-film form of the equations (\ref{thin_film}) for four values of electrowetting number $\epsilon_{wall} = \{ 0.38, 0.46, 0.52, 0.69\}$ (red, blue, yellow, and green lines, respectively) showing the evolution of angle $\theta$ and the contact angle velocity $dX_c/dt$ (insert) as a function of time $t$.}  
\label{Collapse_based_on_electrowet}
\end{figure}

In principle, this approach can be generalized to include the effect of wall interaction using the data from Fig. \ref{Nelson_et_al_2011_vs_MD} (b).
As each electrowetting case has a different equilibrium angle $\theta_e$, we consider the relative change in angle $\langle\theta\rangle - \theta_e$ and, for the four different electrowetting numbers Fig \ref{Nelson_et_al_2011_vs_MD} (b), we can collapse these onto a single curve by multiplying velocity with an arbitrary function of wall interaction $\epsilon_{\textrm{wall}}$, here $\epsilon^{5/2}_{\textrm{wall}}$. 
The data with this scaling is shown in \ref{Collapse_based_on_electrowet} (a).
By taking the coefficient for the line of best fit, $1377.7$, we can define a relationship between angle, sliding velocity and electrowetting, 
\begin{align}
 u = \frac{  \left(\langle\theta\rangle - \theta_e \right)}{ 1377.7 \epsilon_{\textrm{wall}}^{5/3}}
\end{align}
which is simply a form of Tanner's law \eq{Tanners} with $n=1$ and $A$ a function of the electrowetting number.
Using the form of Tanner's law from \eq{Tanners} with the Langevin \eq{Langevin_equations_numerical}, wetting is incorporated into a solver for the continuum thin-film equations \ref{thin_film} as shown in Fig \ref{Collapse_based_on_electrowet} (b) (see \citet{Karapetas_et_al} for implementation details of the thin-film solver and \citet{Smith_et_al2016} for the tuning of the contact line model). 
The equilibrium angles are taken from the receding angles in Fig \ref{Nelson_et_al_2011_vs_MD} (a),  
with $\theta_e = \{112, 98.7, 88.7, 50.5 \}$ for  $\epsilon_{\rm wall} = \{ 0.38, 0.46, 0.52, 0.69 \}$, respectively.
The molecular fluctuations are seen in the contact line velocity of Fig \ref{Collapse_based_on_electrowet} (b insert), with the difference in speed of evolution and final equilibrium angle shown in the main part of Fig \ref{Collapse_based_on_electrowet} (b).
In this way, the molecular detail has been parameterized and included in a form that is directly useful for CFD applications (\ie{} a closure model for contact line motion).

The use of a thin-film solver is potentially invalidated by the large angles present in the MD simulation used to design the model. In practical application, the approach used here should be included as part of a more complex CFD solver modelling larger angles. We apply this technique to the thin-film equations as an example of how molecular contact motion can be parameterized and incorporated into a continuum solver; through Tanner's law using a simple droplet model.
In addition, the use of a reduced model for the contact line represents a massive simplification of very complex molecular detail.
While this \textit{may} be acceptable for flat walls and simple fluids, it would not be expected to work for more complex examples.
These include common challenges in industrial fluid mechanics such as surfactant-laden flow, rough or textured walls, build up of surface fouling, large heat gradients and phase change.

\begin{figure}
        \centering
(a) \hspace{2.8in} (b) \\
        \begin{subfigure}{0.42\textwidth}
		\includegraphics[width=\textwidth]{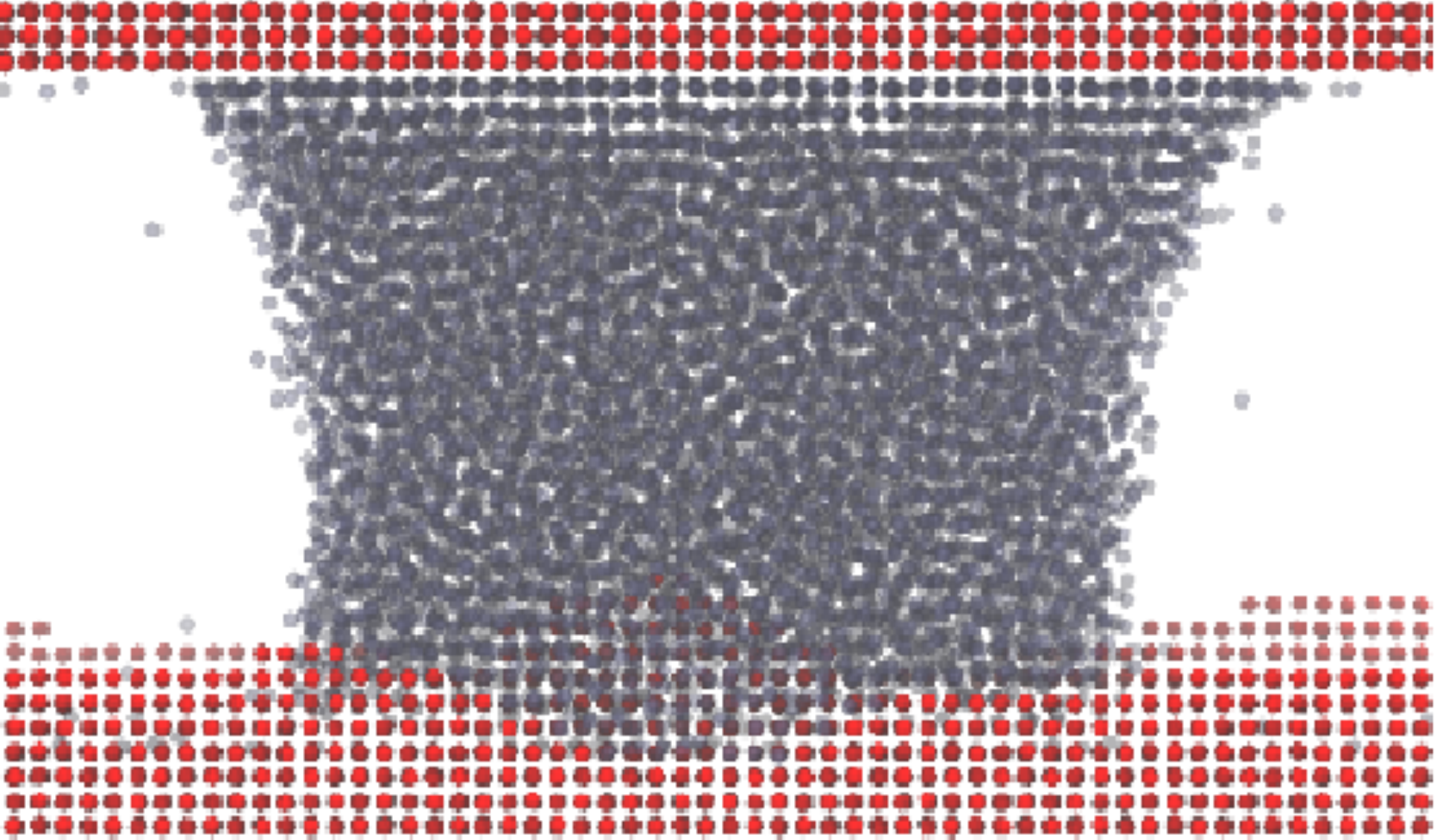}
         \end{subfigure} \;\;
	  \begin{subfigure}[c]{0.44\textwidth}
	  \includegraphics[width=\textwidth]{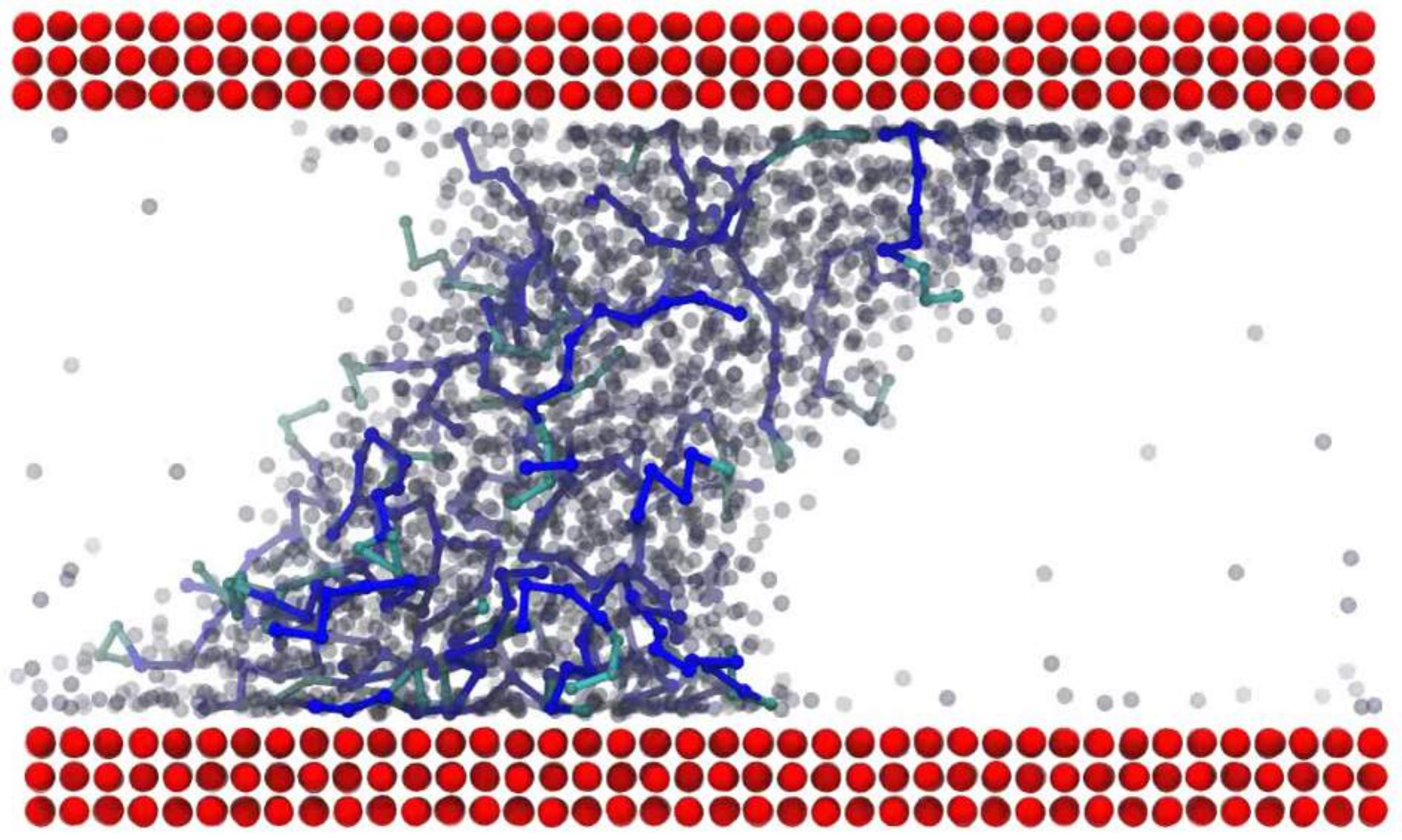}
         \end{subfigure} 
    \caption{(a) Including roughness at the molecular scale using an arbitrary superposition of random cosines to removing molecules from FCC lattice of tethered molecules. 
     (b) Building surfactant molecules into the SAFT-$\gamma$ water sheared liquid model.
}
    \label{fig:rough_wall}
\end{figure}

\begin{figure}
        \centering
	(a) \hspace{3in} (b) \\
        \begin{subfigure}{0.48\textwidth} 
                \includegraphics[width=\textwidth]{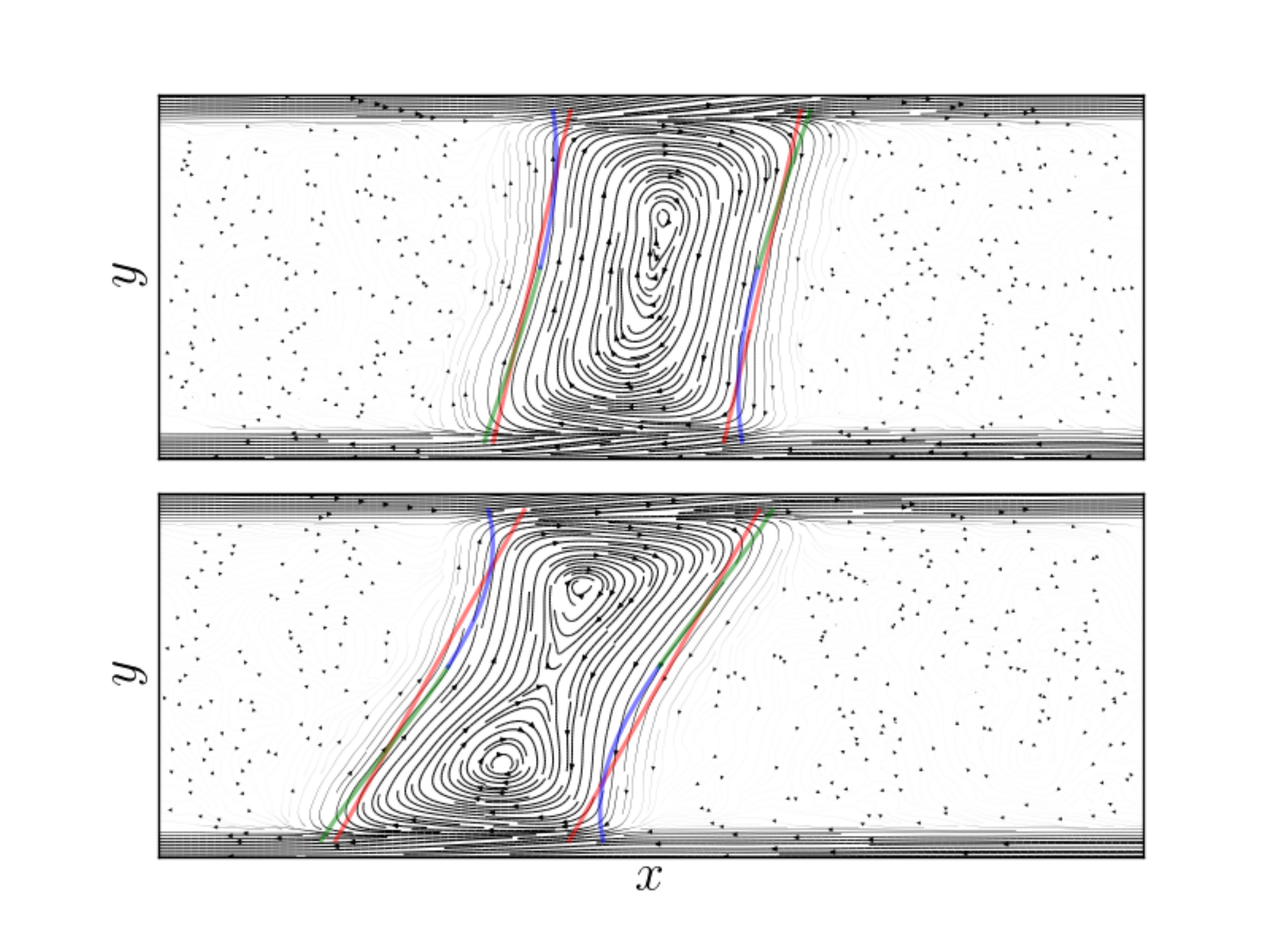}
        \end{subfigure}
        \begin{subfigure}{0.48\textwidth}
                \includegraphics[width=\textwidth]{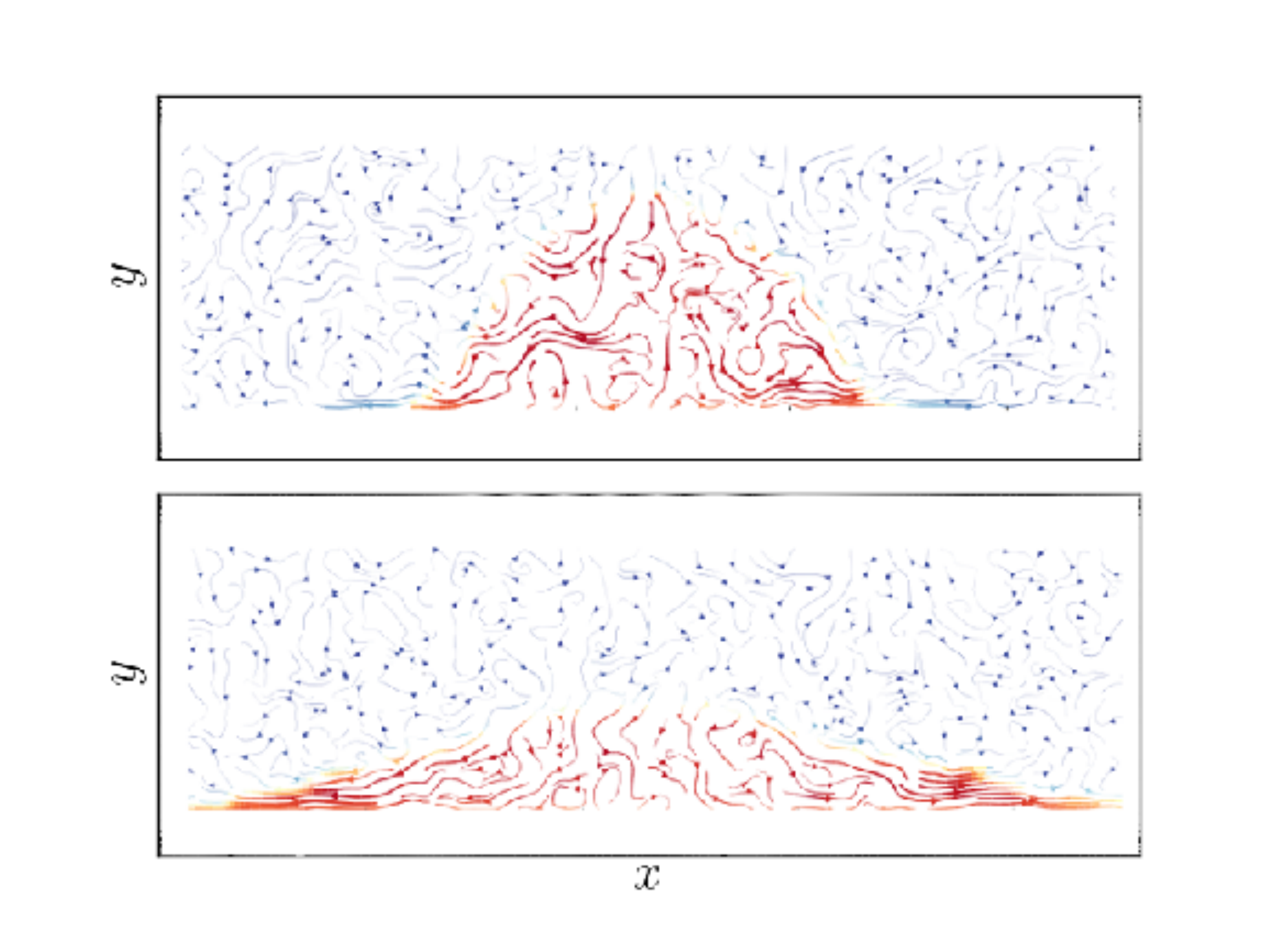}
        \end{subfigure} 
        \caption{A comparison of the streamlines for (a) a liquid bridge at low sliding rates and at the point of pinch off with contact angle highlighted and (b) a droplet before and during spreading with streamlines colored by mean particle density from blue (low) to red (high).}
        \label{Streamlines_droplet_vs_bridge} 
\end{figure}

To address this complexity, one possible future extension could be to use Type 2 coupling, where surfactants are included or fractal wall roughness is modelled explicitly with the contact line dynamics fed back into the continuum model.
Examples of possible embedded MD models are shown in Fig \ref{fig:rough_wall}(a) for fractal roughness and Fig \ref{fig:rough_wall} (b) for inclusion of surfactants.
As the parameter space is too large to model and potentially too complex to define a reduced order model, we could consider a Type 2 embedded scheme, where MD is run as needed to get contact line data at a state point or based on observed roughness of a wall at that point in a CFD model.
However, there are potential issues with this modelling methodology; the flow field may not be representative of an actual droplet and, at higher flow shear rates, the liquid bridge can be seen to pinch off. 
To highlight both the different physics and pinch off, the streamlines for a liquid bridge and a droplet are compared in Fig \ref{Streamlines_droplet_vs_bridge}.
The streamlines observed here are consistent with the variation of flow regimes observed in a continuum study of a similar liquid bridge by \citet{Ren_E_2011}.
The liquid bridge deforms linearly (elastically) for smaller shear rates, returning to a stationary bridge if the shear is removed.
Beyond a certain yield strain, the liquid bridge is pulled apart and fails like a solid in the non-linear (plastic) region.
The streamlines in the liquid bridge of Fig \ref{Streamlines_droplet_vs_bridge} (a) looks similar to a Kirchhoff ellipse vortex, but gradually move apart until the vortex pair becomes sufficiently separated that the middle region (a region of low pressure) is no longer surrounded by a flow and surface tension cannot hold the bridge together.
The pinch off mechanism seen in the molecular system also bears a striking qualitative resemblance to the one observed experimentally \citep{Smith2016, Wang_McCarthy}.
This limitation on the range of stability of a sheared liquid bridge places a constraint on the range of contact line dynamics that can be explored.
This is important for modelling surfactants as reduced surface tension would further promote pinch off.
In addition, when compared to the droplet flow-field shown by the streamlines of Fig \ref{Streamlines_droplet_vs_bridge}(b), it is clear that the fluid dynamics is not the same in the liquid bridge of Fig \ref{Streamlines_droplet_vs_bridge}(a).
This all suggests that the liquid bridge may not be the appropriate method to get the dynamic contact line behavior for Type 2 embedded coupling.
As the flow of surfactant determines contact line motion for superspreading, the next section outlines a model of the entire MD droplet to understand the correct dynamics in the presence of surfactants.
The system size limitations are directly discussed and it is found that large droplets are required to get macroscopic behaviour.

\subsection{Droplet modelling}
\label{sec:droplet}
The molecular-level modelling of surfactant-laden droplets requires the simulation of large systems.\cite{Theodorakis2015b,Santiso2013}
Even in the case of a pure aqueous SAFT-$\gamma$ water droplet without surfactants, for example, the size of the droplet should exceed about $65,000$ effective beads in order to render the contact angle independent of the droplet size \cite{Theodorakis2015b} minimizing line tension effects, which are present in nanoscale droplets.\cite{Weijs2011} 
To this end, line tension was first introduced by Gibbs as `linear tension' suggesting that it might be considered in a manner entirely  analogous to that in which surfaces of discontinuity are treated and it may have negative values, particularly relevant for small systems.\cite{Gibbs1961}

\begin{figure}[H]
\includegraphics[width=\figwidth\textwidth]{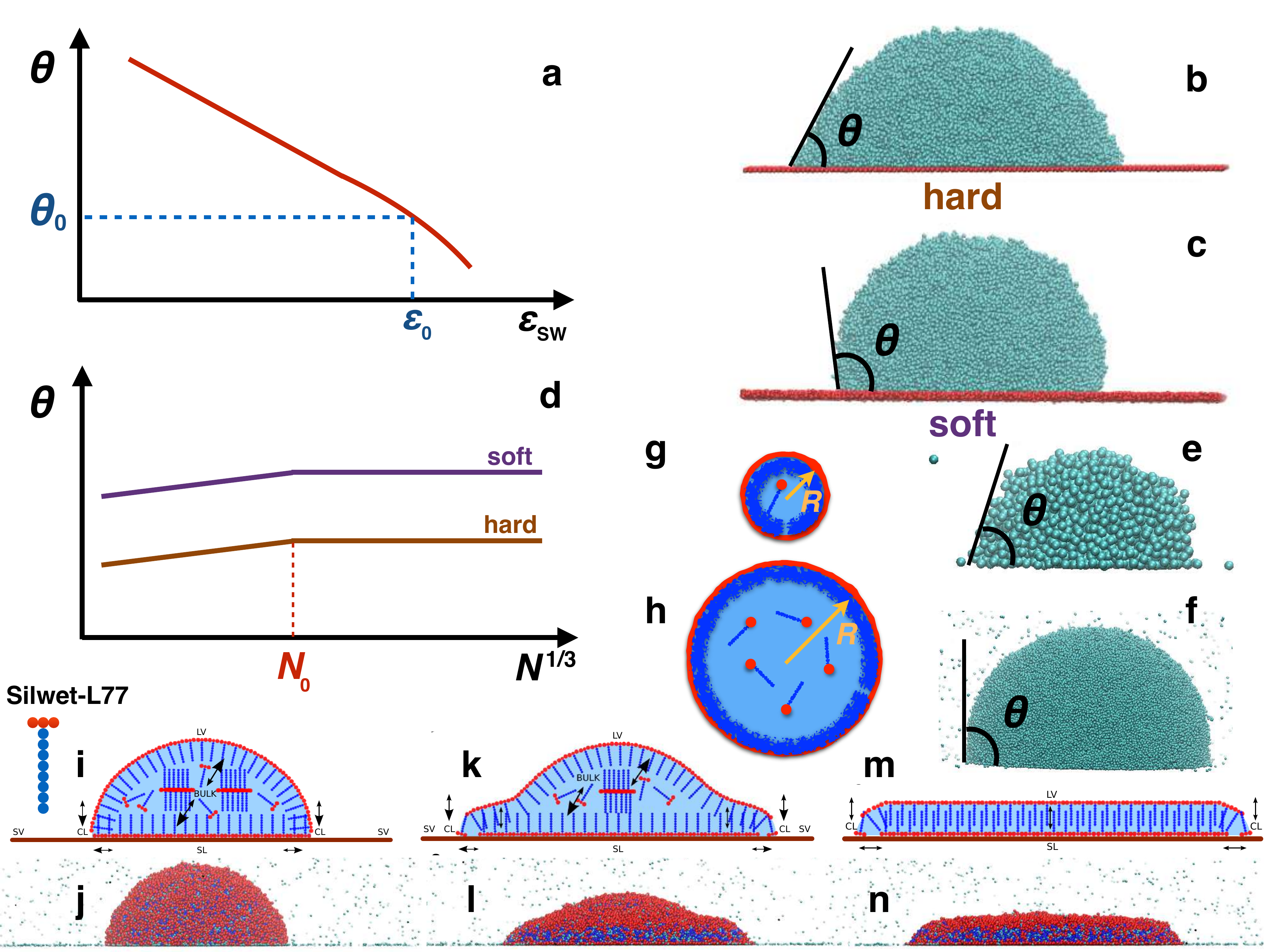}
\caption{(a) The dependence of the contact angle on the potential interaction between the water molecules and an unstructured smooth wall. Contact line phenomena appear for small contact angles below a threshold value $\varepsilon_0$. In the case of the SAFT-$\gamma$ water model this value is about $45^{\rm o}$. The difference in the contact angle of a droplet ``sitting'' on a hard (b) and a soft (c) substrate is illustrated. Panel (d) illustrates the dependence on the size of the droplet expressed through the number of effective beads $N$. The contact angle increases with the size of the droplet up to a threshold value. In the case of the SAFT-$\gamma$ water model this is estimated around $N=65 000$ effective beads.  However, the contact angle is also model dependent. Panels (e) and (f) show example of a small and a large droplet, respectively. Panels (g) and (h) illustrate two spherical droplets of different sizes at the CAC. The droplets have different concentration despite being both at the CAC. 
The CAC concentration scales as $1/R$.\cite{Theodorakis2015} Therefore, CAC (which is different for droplets of different size) provides a unit of concentration that allows for comparison between droplets of different size. Panels (i)-(n) illustrate the characteristic snapshots of a superspreading droplet due to the Silwet-L77 superspreading surfactant. For each snapshot we present a schematic of the main adsorption processes taking place at each state of the superspreading. A schematic of the Silwet-L77 superspreading surfactant in our molecular model is illustrated. Here, blue are beads EO (hydrophilic) and red M and Q beads (hydrophobic). Cyan colored are the SAFT water beads. Adapted from Ref. 12. Copyright 2015 American Chemical Society. }  
\label{droplets_panos}
\end{figure}

An accurate way to measure contact angles by avoiding a fitting procedure is to use the curvature of the droplet through the following relation\cite{Theodorakis2015b},
\begin{equation}
\label{eq:phi}
\theta = \arcsin(1/\mu_{\rm D}),
\end{equation}
where $\mu_{\rm D} = (1+\lambda^2)/(2\lambda)$, $\lambda=h/\alpha$ with $h$ being the distance from the solid--liquid interface to the apex of the droplet and $\alpha$ being the radius of the solid--liquid interface.
The calculation of the contact angle through the ratio $\lambda$ also results in smaller statistical errors as one is only required to measure the ensemble average values of $h$ and $\alpha$. 
After the contact angle has reached a constant value, independent of the droplet size (Fig.~\ref{droplets_panos}), the strength of interaction between the water effective beads and the unstructured smooth substrate can be tuned against experimental data and continuum simulations. \cite{Theodorakis2017,Theodorakis2015b}
However, even above this threshold value, which is also system dependent,
the contact line will still play a significant role when the interaction strength between the droplet and the substrate exceeds a certain value $\varepsilon_0$, which is faster reached on stiff substrates (Fig.~\ref{droplets_panos}) \cite{Theodorakis2017}.

\begin{figure}[H]
\includegraphics[width=8.9cm]{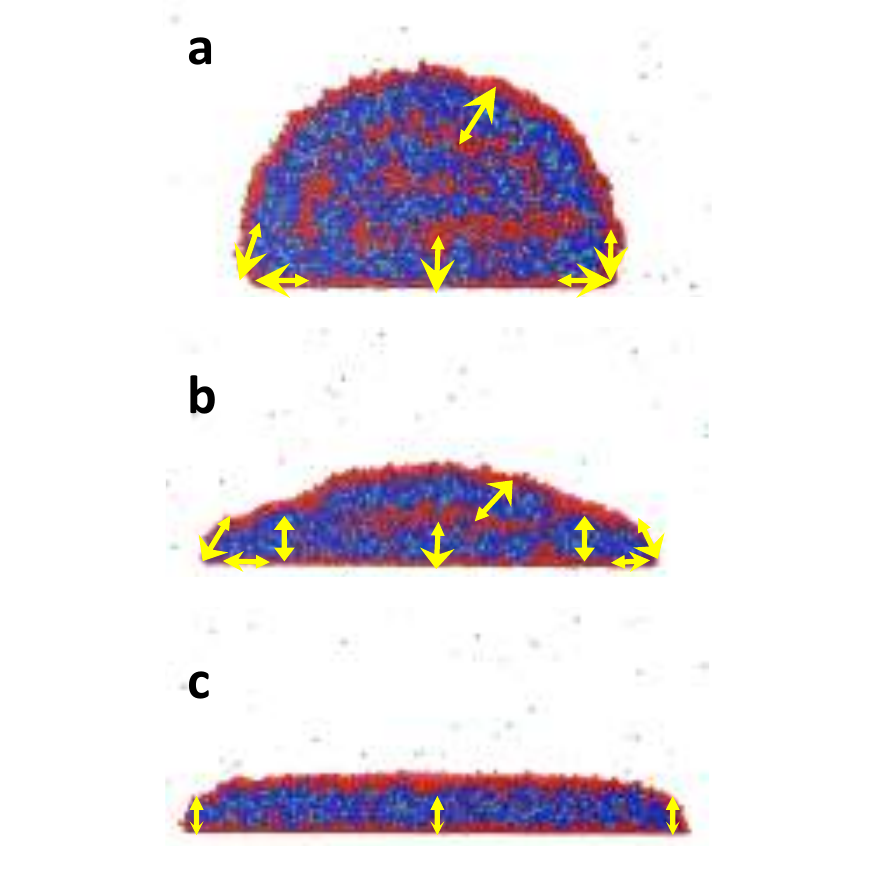}
\caption{Different stages during the spreading of a surfactant-laden droplet (the superspreader Silwet-L77 surfactant has been used here and the cross-section of the droplet is illustrated): an initial (a) and an intermediate (b) non-equilibrium states, and a final equilibrium
state (c). The main adsorption mechanisms at each stage of the spreading process
are indicated with differences between the arrow heads indicating the dominant direction of the adsorption process during
spreading. Red color indicates M and Q hydrophobic beads, whereas blue indicates EO hydrophilic groups. Water molecules are in cyan color.}  
\label{mechanisms_panos}
\end{figure}

While we can overcome the contact angle dependence on the droplet size by reaching system sizes of the order of $10^5$ effective beads, systems with surfactants pose a much greater challenge for simulations. 
In this case, one needs to reach a macroscopic limit to capture the various processes that take place within the droplet, for example, the diffusion of surfactant monomers/aggregates within the droplet or 
the formation/dissociation of aggregates. 
In fact, the spreading mechanism also depends on the molecular shape of the liquid molecules \cite{Shanahan1998,Wu2017} and the surfactant affecting the contact line motion,\cite{Song2017} where the latter has been investigated by MD simulations in the context of superspreading.\cite{Theodorakis2015} 
We should note here that the critical aggregation concentration (CAC) for nanoscale droplets depends strongly on the size of the droplet as well.
The surfactant concentration, $w$, which is usually expressed 
through wt\%, scales as $1/R$ for spherical droplets (prefactors have been omitted here), where $R$ is the radius of the droplet.
This simply means that the absolute values of concentration do not provide a measure for direct comparison between droplets of different sizes and concentrations may be better expressed as a multiple 
of the CAC concentration for each droplet. 
Although from the point of view of simulation experiments the ideal
situation would be to simulate macroscopic droplets where the above problems gradually disappear, one can still identify certain adsorption mechanisms for nanoscale droplets as illustrated in Fig.~\ref{mechanisms_panos} for the superspreading of Silwet-L77 surfactant.\cite{Theodorakis2015,Theodorakis2015b}

%

In the superspreading example (Fig.~\ref{mechanisms_panos}), MD simulations can provide information for the dynamics of the contact line for Type 1 and 2 coupling. 
However, there is still the need to be able to model even larger systems in many cases.
This is very important in the context of Fig. \ref{mechanisms_panos}, where different processes take place within the droplet.
Although adsorption processes characteristic of the superspreading, such as the adsorption of surfactant onto the substrate through the contact line or the replenishment of surfactant at the liquid--vapour interface, can be described by MD simulations and provide information for Type 1 and 2 coupling, still, diffusion processes or dynamic formation of aggregates cannot be handled by MD without using excessively large system sizes (Fig.~\ref{mechanisms_panos}).
All this emphasizes the challenges of MD simulations to tackle some of these issues, which renders Type 3 coupling an obvious solution for realizing the full simulation of the droplet. 
As Type 3 coupling is limited to the molecular time and length scales, the future of coupling will likely use the output from these simulations to either inform Type 1 or even provide results for a Type 2 coupling.
As the field advances, solutions mixing the advantages of the various approaches will likely become more routine.

\section{Conclusions}

Molecular simulation has shown great promise in getting \textit{a priori} results for near-wall behaviour, liquid--vapour interfaces and dynamics of complex molecules such as surfactants. In the latter case, the molecular architecture of the liquid molecules and the surfactants is of great importance as it affects significantly the spreading mechanisms of a droplet.\cite{Shanahan1998,Wu2017,Song2017} 
Moreover, for submicron/nano-sized droplets, the surface tension and three-phase contact angle are a function of the drop size.\cite{Ono_Kondo}
The moving contact line has to incorporate all this complexity and furthermore couple this with slip, surface tension and complex bulk--fluid flow.
In this invited article, the anatomy of the moving contact line is explored by analyzing the effect of these various contributions, gradually building up to the full complexity of the dynamic contact line on a superspreading droplet.
In doing this, we explore techniques which can be used to couple the molecular model to continuum simulation, both directly as part of the same simulation and indirectly by parameterizing equations.

To understand the key quantities MD could give the continuum, we study each of the contact line problems constituent parts: slip-length, vapour--liquid surface tension, static contact angle behaviour and contact line dynamics.
Starting with slip, the near-wall effects are explored for a single phase.
A direct coupling approach is presented here, retaining a region of molecular detail and linking the two systems on an interface.
Next, the liquid--vapour interface is analyzed through the surface tension and the effect of surfactants is explored.
The static angle in a liquid--vapour system is then parameterized for a range of wall interactions potentials by comparison to experiments.
This is then extended to include sliding of these walls to understand how these behave when the contact angle is moving. 
The molecular sliding is compared to experiments, noting broadly similar behaviour despite the difference in scale between the two systems but ultimately poor agreement, attributed to the limited simulation sizes possible with MD. 
In order to couple to CFD, a simplified contact line model parameterized using MD is presented.
The limitations of these reduced models for the contact line are discussed in detail and the work finishes by presenting a full large-scale simulation of a molecular droplet with surfactants.

We conclude  by noting that, although clearly promising, the methodology of coupling MD to continuum models to capture the precise dynamics of the contact line for a range of non-trivial situations, for instance, the presence of surfactants, surface wettability and chemical reactions, is at an early stage.
The use of coupling relies on the validity of molecular simulation, which is difficult to compare with experiments given the scale separation. However, as computers get bigger and experiments higher resolution, direct comparisons of the two approaches is increasingly possible. Coupled simulation provides an opportunity to accelerate the comparison by allowing simulations of larger scales.
We have outlined current strategies, noting that the future of MD/continuum coupling will likely combine a range of these approaches, and hope that it encourages and motivates others to engage in further research into this area.

\subsubsection{Acknowledgements}
This work is supported by the EPSRC Platform Grant MACIPh (Grant number EP/L020564/1). This research has been supported by the National Science Centre, Poland, under grant No.~2015/19/P/ST3/03541. This project has received funding from the European Union's Horizon 2020 research and innovation programme under the Marie Sk{\l}odowska--Curie grant agreement No. 665778. This research was supported in part by PLGrid Infrastructure.

\bibliography{./ref}

\newpage
\section{Graphical TOC Entry}
\begin{figure}
\centering
\fbox{\includegraphics[width=3.2in]{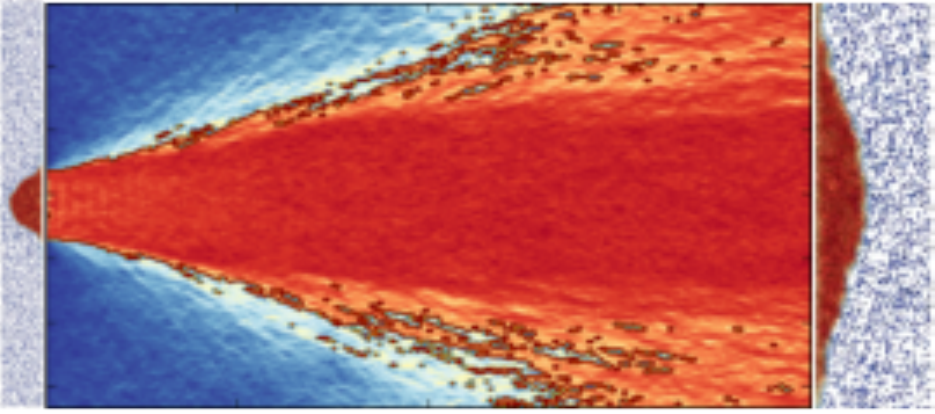}}
\end{figure}

\end{document}